\documentclass[runningheads]{llncs}
\usepackage{graphicx}
\newif\ifTR
\TRtrue

\ifTR
\newcommand{\refappendix}[1]{\reftxt{Appendix}{#1}}
\else
\newcommand{\refappendix}[1]{the extended version~\cite{brbotr}}
\fi

\usepackage[lncs]{bectex/bec} 
\usepackage{commons}
\begin{document}
\ifTR
\title{Selectively-Amortized Resource Bounding (Extended Version)}
\else
\title{Selectively-Amortized Resource Bounding}
\fi
%
%
\author{Tianhan Lu, Bor-Yuh Evan Chang, Ashutosh Trivedi}
\authorrunning{Lu et al.}
\institute{University of Colorado Boulder\\
\email{\{tianhan.lu,bec,ashutosh.trivedi\}@colorado.edu}}
%
%
\maketitle              
\begin{abstract}
  We consider the problem of automatically proving resource bounds.
  That is, we study how to prove that an integer-valued resource variable is
  bounded by a given program expression. 
  Automatic resource-bound analysis has recently received significant
  attention because of a number of important applications (e.g., detecting
  performance bugs, preventing algorithmic-complexity attacks, identifying
  side-channel vulnerabilities), where the focus has often been on developing
  precise amortized reasoning techniques to infer the most exact resource
  usage. 
  While such innovations remain critical, we observe that fully precise
  amortization is not always necessary to prove a bound of interest.
  And in fact, by amortizing \emph{selectively}, the needed supporting
  invariants can be simpler, 
  making the invariant inference task more feasible and predictable.
  We present a framework for selectively-amortized analysis that
  mixes worst-case and amortized reasoning via a property decomposition and a
  program transformation. 
  We show that proving bounds in any such decomposition yields a sound resource
  bound in the original program, and we give an algorithm for selecting a
  reasonable decomposition. 
\end{abstract}

\begin{center}
  \includegraphics[scale=0.16]{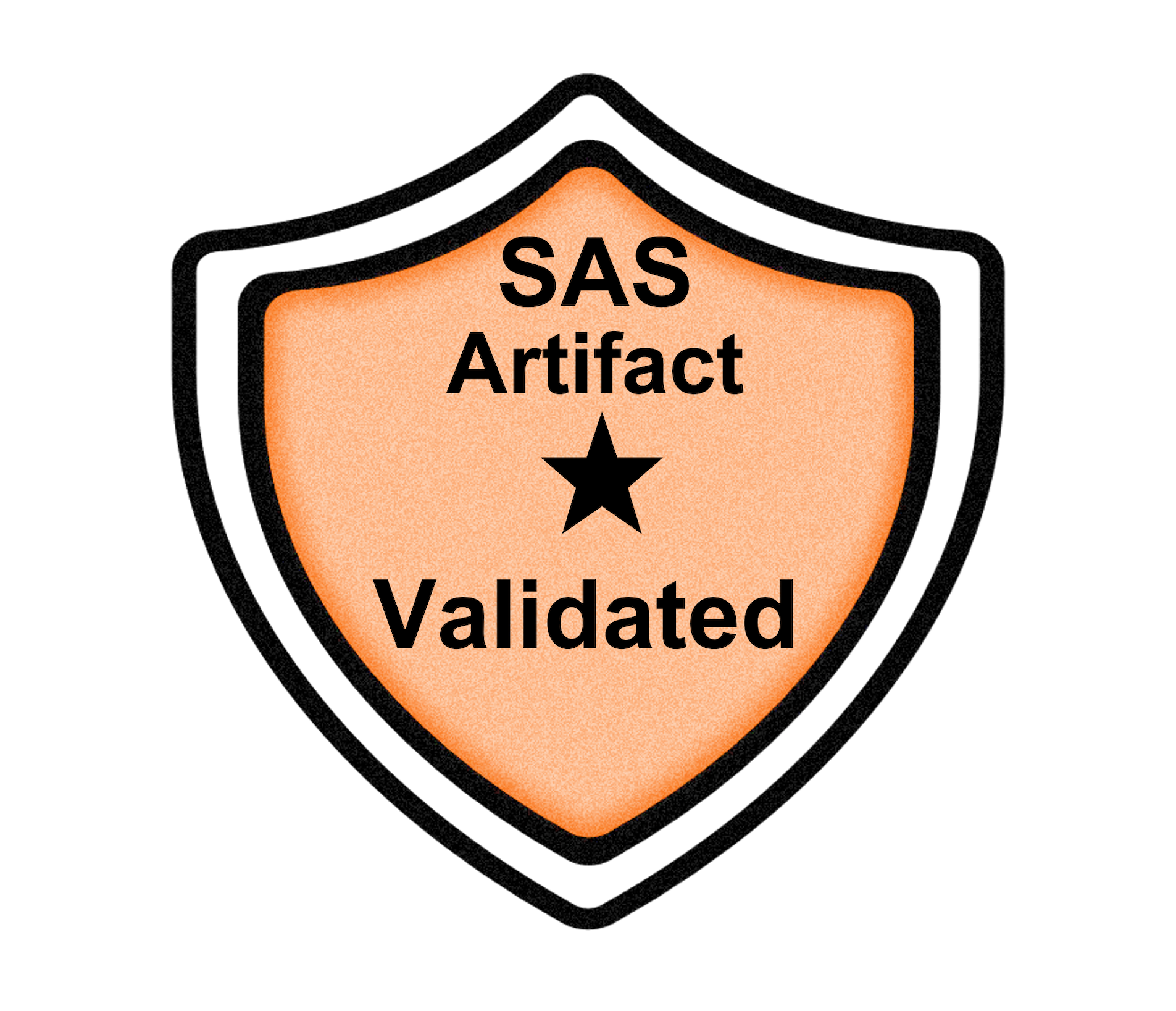} \includegraphics[scale=0.16]{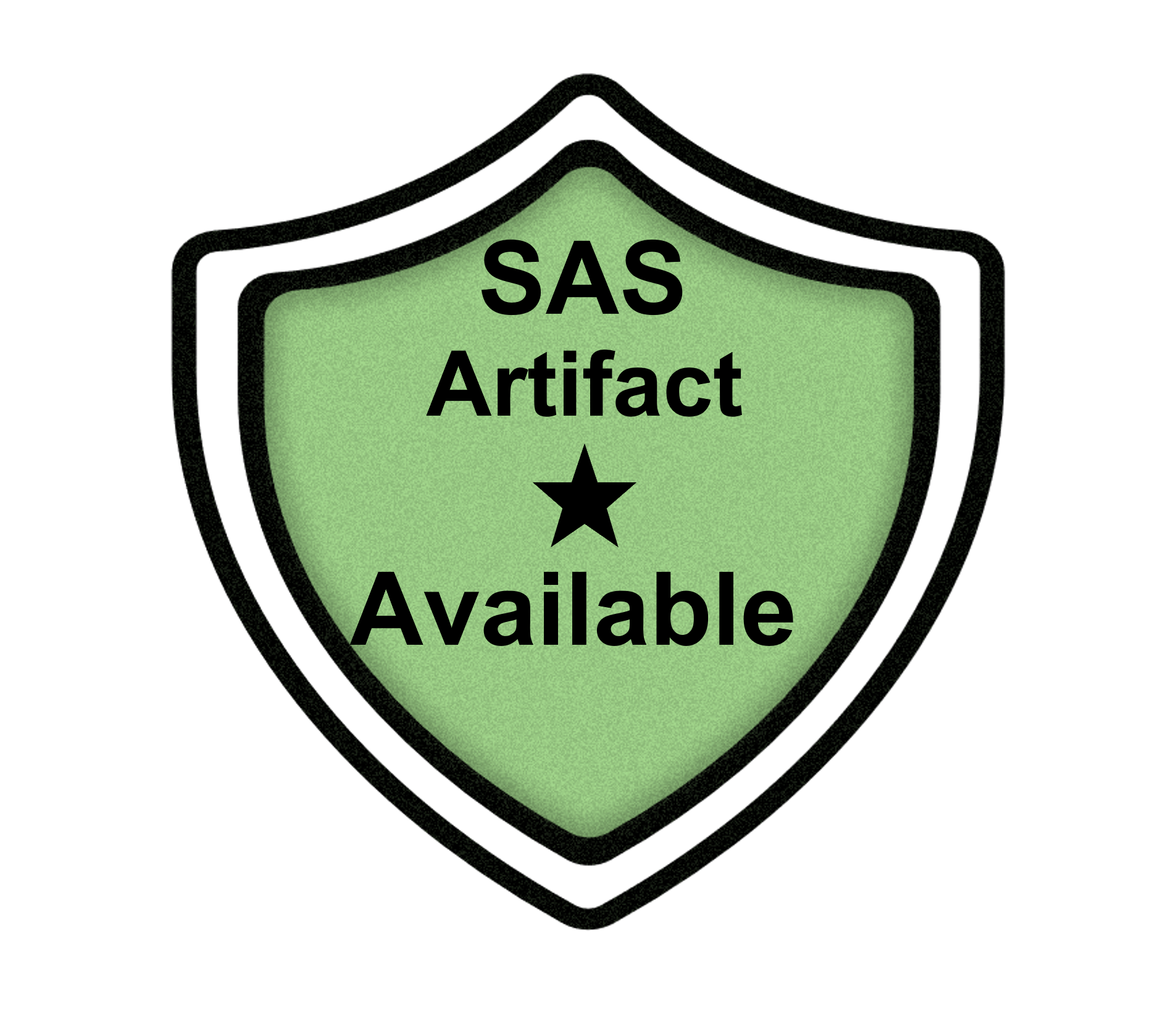}%
\end{center}

\newcommand{\toolname}{\hbox{\textsc{Brbo}}\xspace}

\section{Introduction}
\label{sec:introduction}
%
In recent years, automatic resource-bound analysis has become an increasingly
specialized area of automated reasoning because of a number of important and 
challenging applications, including statically detecting performance bugs,
preventing algorithmic-complexity attacks, and identifying side-channel
vulnerabilities. 
In this paper, we consider the specific problem of proving bounds on
resource usage as follows: given an integer-valued resource variable $\resvar$
that models resource allocation and deallocation, prove that it is
bounded by an expression $\bndsym$ at any program location---that is, prove $\assert{\resvar \le \bndsym}$ anywhere in the program.
Resource allocations and deallocations can be modeled by (ghost) updates
$\auxupdateusecolor{\resvar}{\useexpr}$  to the resource variable $\resvar$ (expressing
that resource usage captured by $\resvar$ increments by $e$ units), and we
generically permit updates to be any expression $\useexpr$.
For example, resource variables can model lengths of
dynamically-sized collections like lists and strings (e.g., \code{List.size()} or
\code{StringBuilder.length()} in Java), and resource updates capture growing or
shrinking such collections (e.g., \code{List.add(Object)},
\code{List.remove(Object)}, or \code{StringBuilder.append(String)}). 


There are two natural ways to address this problem, by analogy to amortized
computational complexity~\cite{tarjan1985amortized}, for which we give intuition
here. 
The first approach views the problem as an extension of the loop bounding
problem, that is, inferring an upper bound on the number of times a loop
executes~\cite{DBLP:conf/fmcad/SinnZV15,DBLP:journals/jar/SinnZV17,DBLP:conf/tacas/BrockschmidtEFFG14,DBLP:conf/pldi/GulwaniJK09,DBLP:conf/pldi/GulwaniZ10,DBLP:conf/sas/ZulegerGSV11,DBLP:conf/cav/SinnZV14}. 
Then to derive upper bounds on resource variables $\resvar$, multiply the worst-case, or upper bound, of an update expression $\useexpr$ by an upper bound on the number of times that update is executed, summed over each resource-use command $\auxupdateusecolor{\resvar}{\useexpr}$, thereby leveraging the existing machinery of loop bound analysis~\cite{DBLP:journals/jar/SinnZV17,DBLP:conf/tacas/BrockschmidtEFFG14,DBLP:conf/fmcad/CadekDSZ18}.
We call this approach \emph{worst-case reasoning}, as it considers the worst-case cost of a given resource-use command for each loop iteration.
This worst-case reasoning approach has two potential drawbacks.
First, it presupposes the existence of loop bounds (\ie assumes terminating programs), whereas we may wish to prove resource usage remains bounded in non-terminating, reactive programs (\eg~\citet{DBLP:conf/vmcai/LuCC019}) or simply where loop bounds are particularly challenging to derive. Second, as the terminology implies, it can be overly pessimistic because the value of the resource-use expression $\useexpr$ may vary across loop iterations.

The second approach to resource bound verification 
is to directly adopt the well-established method of finding inductive invariants
strong enough to prove assertions~\cite{DBLP:journals/tcs/MannaP91}. 
However, directly applying inductive invariant inference techniques (e.g.,
\citet{DBLP:conf/cav/SharmaDDA11,DBLP:conf/oopsla/DilligDLM13,DBLP:conf/cav/ColonSS03,DBLP:conf/pldi/KincaidBBR17,DBLP:journals/pacmpl/KincaidCBR18,DBLP:journals/pacmpl/KincaidBCR19,DBLP:conf/lics/HrushovskiOP018,DBLP:conf/pldi/Chatterjee0GG20})
to the resource bounding can be challenging, because the required
inductive invariants are often particularly complex (e.g., polynomial) and are
thus not always feasible or predictable to infer
automatically~\cite{DBLP:conf/popl/GulwaniMC09,DBLP:conf/fmcad/CadekDSZ18}. 
We call this approach \emph{fully-amortized reasoning}, as the strongest
inductive invariant bounding the resource variable $\resvar$ may consider
arbitrary relations to reason about how the resource-use expression $\useexpr$
may vary across loop iterations, thereby reasoning about amortized costs across
loop iterations.

The key insight of this paper is that the choice is not binary but rather the above two approaches are extremal instances on a spectrum of \emph{selective amortization}.
We can apply amortized reasoning within any sequence of resource updates and then reason about each sequence's contribution to the overall resource usage with worst-case reasoning.
We show that the \emph{decomposition} of the overall resource usage into \emph{amortized segments} can be arbitrary, so it can be flexibly chosen to simplify inductive invariant inference for amortized reasoning of resources or to leverage loop bound inference where it is possible, easy, and precise.
We then realize this insight through a program transformation that expresses a particular class of decompositions and enables using off-the-shelf amortized reasoning engines.
In particular, we make the following contributions:
\begin{enumerate}

  \item We define a space of amortized reasoning based on decomposing resource
    updates in different ways and then amortizing resource usage within the
    resulting segments (\secref{decomposing}). 
    Different decompositions \emph{select} different amortizations, and we prove
    that any decomposition yields a sound upper bound. 
  

  \item We instantiate selective amortization through a program transformation for a particular class of decompositions and define a notion of \emph{non-interfering amortization segments} to suggest a segmentation strategy (\secref{selecting-decomposition}).



  

  \item We implemented a proof-of-concept of selective amortization in a tool \toolname (for \emph{br}eak-and-\emph{bo}und) that selects a decomposition and then delegates to an off-the-shelf invariant generator for amortized reasoning (\secref{experiment}).
  Our empirical evaluation provides evidence that selective amortization effectively leverages both worst-case and amortized reasoning.
\end{enumerate}


Our approach is agnostic to the underlying amortized reasoning engine.
Directly applying a relational inductive invariant generator on resource
variables, as we do in our proof-of-concept (\secref{experiment}), corresponds
to an aggregate amortized analysis, however this work opens opportunities to consider
other engines based 
on alternative amortized reasoning (e.g., the potential
method~\cite{DBLP:conf/popl/HoffmannAH11,DBLP:conf/popl/HoffmannDW17}). 

\newsavebox{\SBoxSBSize}\savebox{\SBoxSBSize}{\code{#sb}}
\newsavebox{\SBoxSepSize}\savebox{\SBoxSepSize}{\code{#sep}}

\section{Overview}
\label{sec:overview}

\autoref{fig:replaceTags-example} shows the core of Java template engine class from the DARPA STAC~\cite{stac} benchmarks.
The \code{replaceTags} method applies a list of templates \code{ts} to the input \code{text} using an intermediate \code{StringBuilder} resource \code{sb} that we wish to bound globally.
In this section, we aim to show that proving such a bound on \code{sb} motivates selective amortized reasoning.

\begin{figure}[tb]
$\lststopn$\begin{lstlisting}[language=Java, alsolanguage=Brbo, xleftmargin=1em]
private String text;
private List<Pair<Integer,Integer>> tags = $\ldots$text$\ldots$;
public String replaceTags(List<Templated> ts, String sep) {$\lstbeginn$
  @Bound(#sb$\;\leq\;$#ts$\cdot($#text$\,+\,$#tags$\cdot$ts#rep$\,+\,$#sep$)$) StringBuilder$\label{line:bound-assertion}\lststopn$
  sb = new StringBuilder();$\label{pt:sb-alloc}\lststartn$
  for (Templated t : ts) {$\label{pt:replaceTags-outerloop}$
    int p = 0;
    for (Pair<Integer,Integer> lr : tags) {$\label{pt:replaceTags-innerloop}$
      int l = lr.getLeft(); int r = lr.getRight();$\label{pt:assume-ordered-tags}\lststopn$
        assume(p $\leq$ l $\leq$ r $\leq$ #text);$\lststartn$
      sb.append(text.substring(p, l));$\label{pt:append-before-tag}$
      String rep = $\ldots$t$\ldots$lr$\ldots$; assume(#rep $\leq$ ts#rep);$\label{pt:assume-rep-bound}$
      sb.append(rep);$\label{pt:append-rep}$
      p = r;
    }
    sb.append(text.substring(p, text.length()));$\label{pt:append-text-leftover}$
    sb.append(sep);$\label{pt:append-separator}$
  }$\lststopn$
  return sb.toString();
}$\lststartn$
\end{lstlisting}
\caption{Motivating selective worst-case and amortized reasoning to analyze a Java template engine class (\texttt{com.cyberpointllc.stac.template.TemplateEngine}).
An instance of this class stores some \code{text} that may have tags in it to replace with this engine.
The tag locations are stored as an ordered list of pairs of start-end indexes in the \code{tags} field, which is computed from \code{text}.
Suppose we want to globally bound the size of the \code{StringBulder}\;\code{sb} used by the \code{replaceTags} method to apply a list of templates \code{ts}.
Let \usebox{\SBoxSBSize} be a resource variable modeling the length of \code{sb} (\ie ghost state that should be equal to the run-time value of \code{sb.length()}).
We express a global bound on \usebox{\SBoxSBSize} to prove with the \code{@Bound} annotation---here in terms of resource variables on the inputs \code{ts}, \code{text}, \code{tags}, and \code{sep}.
}
\label{fig:replaceTags-example}
\end{figure}

At a high-level, the \code{replaceTags} method allocates a fresh \code{StringBuilder}\;\code{sb} to copy non-tag text or to replace tags using the input templates \code{ts} from the input \code{text}.
The inner loop at \ptref{replaceTags-innerloop} does this copy or tag replacement by walking through the ordered list of tag locations \code{tags} to copy the successive ``chunks'' of non-tag text \code{text.substring(p, l)} and a tag replacement \code{rep} at program points~\ref{pt:append-before-tag} and~\ref{pt:append-rep}, respectively
(the \code{assume} statement at \ptref{assume-ordered-tags} captures the ordered list of locations property).
Then, the leftover text \code{text.substring(p, text.length())} after the last tag is copied at \ptref{append-text-leftover}.
The outer loop at \ptref{replaceTags-outerloop} simply does this template-based tag replacement, and inserts a separator \code{sep} (at \ptref{append-separator}), for each template \code{t}.
There are four program points where resources of interest are \emph{used} (\ie \code{sb} grows in length)---the \code{sb.append($\ldots$)} call sites mentioned here.

\newsavebox{\SBoxBoundAssertion}
\begin{lrbox}{\SBoxBoundAssertion}\lstinline[language=Brbo]~|#sb$\;\leq\;$#ts$\cdot($#text$\,+\,$#tags$\cdot$ts#rep$\,+\,$#sep$)$|~\end{lrbox}

The \code{|@Bound|} assertion shown on \lineref{bound-assertion}
\[
  \usebox{\SBoxBoundAssertion}
\]
follows the structure of the code sketched above.
The template-based tag replacement is done \code{|#ts|} number of times where \code{|#ts|} models the size of the template list \code{ts}.
Then, the length of the tag-replaced text is bounded by the length of \code{text} (\ie \code{|#text|}) plus a bound on the length of all tag-replaced text \lstinline[language=Brbo]~|#tags$\,\cdot\,$ts#rep|~ plus the length of the separator \code{sep} (\ie \code{|#sep|}).
A bound on each tag replacement \code{rep} is modeled with a variable \code{|ts#rep|} (which we name with \code{ts} to indicate its correspondence to a bound on all tag replacements described by input \code{ts}) and the \lstinline[language=Brbo]~assume(|#rep$\;\leq\;$ts#rep|)~ statement at \ptref{assume-rep-bound}.
Thus, a bound on the length of all tag-replaced text is \lstinline[language=Brbo]~|#tags$\,\cdot\,$ts#rep|~.
Note that the coloring here is intended to ease tracking key variables but having color is not strictly necessary for following the discussion.

For explanatory purposes, the particular structure of this bound assertion also suggests a mix of worst-case and amortized reasoning that ultimately leads to our selectively-amortized reasoning approach that we describe further below.
Starting from reasoning about the inner loop, to prove that the copying of successive ``chunks'' of \code{text} is bounded by \code{|#text|} requires amortized reasoning because the length of \code{text.substring(p, l)} at \ptref{append-before-tag} varies on each loop iteration.
In contrast, we bound the length of all tag-replaced text with \lstinline[language=Brbo]~|#tags$\,\cdot\,$ts#rep|~ using worst-case reasoning: we assume a worst-case bound on the length of replacement text \code{rep} is \code{|ts#rep|}, so a worst-case bound with \code{|#tags|} number of tag replacements is \lstinline[language=Brbo]~|#tags$\,\cdot\,$ts#rep|~.
Now thinking about the rest of the body of the outer loop at \ptref{append-text-leftover}, the leftover text copy is amortized with the inner loop's copying of successive ``chunks,'' so we just add the length of the separator \code{|#sep|}.
Finally, considering the outer loop, we simply consider this resource usage bound for each loop iteration to bound \code{|#sb|} with \lstinline[language=Brbo]~|#ts$\cdot(\ldots)$|~.

The key observation here is that to prove this overall bound on \code{|#sb|},
even though we need to amortize the calls to \code{sb.append(text.substring(p,
l))} at \ptref{append-before-tag} over the iterations of the inner loop, we do
not need to amortize the calls at this same site across iterations of the outer
loop.
Next, we translate this intuition into an approach for \emph{selectively-amortized resource bounding}.

\newsavebox{\SBoxLMinusPOp}\savebox{\SBoxLMinusPOp}{\code{|op:(|l |op:-| p|op:)|}}
\newsavebox{\SBoxTextMinusPOp}\savebox{\SBoxTextMinusPOp}{\code{|op:(|#text |op:-| p|op:)|}}

\newsavebox{\SBoxBound}
\begin{lrbox}{\SBoxBound}\lstinline[language=Brbo]~#sb$\;\leq\;$#ts$\cdot($#text$\,+\,$#tags$\cdot$ts#rep$\,+\,$#sep$)$~\end{lrbox}

\newsavebox{\SBoxBoundRHS}
\begin{lrbox}{\SBoxBoundRHS}\lstinline[language=Brbo]~#ts$\cdot($#text$\,+\,$#tags$\cdot$ts#rep$\,+\,$#sep$)$~\end{lrbox}

\subsection{Decomposing Resource Updates to Selectively Amortize}
\label{sec:overview-decomposing}

The resource-bound reasoning from \figref{replaceTags-example} may be similarly expressed in a numerical abstraction where all variables are of integer type as shown in \figref{inv-original}.
There, we write $\auxupdateusecolor{\resvar}{\useexpr}$ for tracking $\useexpr$ units of resource use in $\resvar$ and $\assign{\var}{\ast}$ for a havoc (\ie a non-deterministic assignment).
Note that \code{text.substring(p, l)} translates to \usebox{\SBoxLMinusPOp}.
To express checking the global bound, we write \code{assert(#sb $\;\le\;$ $\bndsym$)} after each $\auxupdatekwusecolor$ update. We also note a pre-condition that simply says that all of the inputs sizes are non-negative.
Crucially, observe to precisely reason about the resource usage \code{#sb} across all of these updates to \code{#sb} requires a polynomial loop invariant, as shown at \ptref{na-complex-inv} in braces \code{\{}$\cdots$\code{\}}.

\newsavebox{\SBoxBoundInvExample}
\begin{lrbox}{\SBoxBoundInvExample}
\begin{lstlisting}[language=Java,alsolanguage=Brbo,numbers=none]
globally($\usebox{\SBoxBound}$);
\end{lstlisting}
\end{lrbox}


\newsavebox{\SBoxSBUpdateOne}
\begin{lrbox}{\SBoxSBUpdateOne}\lstinline[language=Brbo]~$\auxupdateusecolor{\usebox{\SBoxSBSize}}{\usebox{\SBoxLMinusPOp}}$;~\end{lrbox}
\newsavebox{\SBoxSBUpdateTwo}
\begin{lrbox}{\SBoxSBUpdateTwo}\lstinline[language=Brbo]~$\auxupdateusecolor{\usebox{\SBoxSBSize}}{\texttt{\#rep}}$;~\end{lrbox}
\newsavebox{\SBoxSBUpdateThree}
\begin{lrbox}{\SBoxSBUpdateThree}\lstinline[language=Brbo]~$\auxupdateusecolor{\usebox{\SBoxSBSize}}{\usebox{\SBoxTextMinusPOp}}$;~\end{lrbox}
\newsavebox{\SBoxSBUpdateFour}
\begin{lrbox}{\SBoxSBUpdateFour}\lstinline[language=Brbo]~$\auxupdateusecolor{\usebox{\SBoxSBSize}}{\usebox{\SBoxSepSize}}$;~\end{lrbox}

\newsavebox{\SBoxSBOne}
\begin{lrbox}{\SBoxSBOne}\lstinline[language=Brbo]~$\clrone{\texttt{\#sb1}}$~\end{lrbox}
\newsavebox{\SBoxSBTwo}
\begin{lrbox}{\SBoxSBTwo}\lstinline[language=Brbo]~$\clrtwo{\texttt{\#sb2}}$~\end{lrbox}
\newsavebox{\SBoxSBThree}
\begin{lrbox}{\SBoxSBThree}\lstinline[language=Brbo]~$\clrthree{\texttt{\#sb3}}$~\end{lrbox}

\newsavebox{\SBoxSBOneReset}
\begin{lrbox}{\SBoxSBOneReset}\lstinline[language=Brbo]~$\resetusecolor{\usebox{\SBoxSBOne}}$~\end{lrbox}
\newsavebox{\SBoxSBTwoReset}
\begin{lrbox}{\SBoxSBTwoReset}\lstinline[language=Brbo]~$\resetusecolor{\usebox{\SBoxSBTwo}}$;~\end{lrbox}
\newsavebox{\SBoxSBThreeReset}
\begin{lrbox}{\SBoxSBThreeReset}\lstinline[language=Brbo]~$\resetusecolor{\usebox{\SBoxSBThree}}$;~\end{lrbox}

\newsavebox{\SBoxSBOneUpdateOne}
\begin{lrbox}{\SBoxSBOneUpdateOne}\lstinline[language=Brbo]~$\auxupdateusecolor{\usebox{\SBoxSBOne}}{\usebox{\SBoxLMinusPOp}}$;~\end{lrbox}
\newsavebox{\SBoxSBOneUpdateTwo}
\begin{lrbox}{\SBoxSBOneUpdateTwo}\lstinline[language=Brbo]~$\auxupdateusecolor{\usebox{\SBoxSBOne}}{\usebox{\SBoxTextMinusPOp}}$;~\end{lrbox}
\newsavebox{\SBoxSBTwoUpdate}
\begin{lrbox}{\SBoxSBTwoUpdate}\lstinline[language=Brbo]~$\auxupdateusecolor{\usebox{\SBoxSBTwo}}{\texttt{\#rep}}$;~\end{lrbox}
\newsavebox{\SBoxSBThreeUpdate}
\begin{lrbox}{\SBoxSBThreeUpdate}\lstinline[language=Brbo]~$\auxupdateusecolor{\usebox{\SBoxSBThree}}{\usebox{\SBoxSepSize}}$;~\end{lrbox}

\newsavebox{\SBoxSBOneCount}
\begin{lrbox}{\SBoxSBOneCount}\lstinline[language=Brbo]~$\clrone{\cnt{\usebox{\SBoxSBOne}}}$~\end{lrbox}
\newsavebox{\SBoxSBTwoCount}
\begin{lrbox}{\SBoxSBTwoCount}\lstinline[language=Brbo]~$\clrtwo{\cnt{\usebox{\SBoxSBTwo}}}$~\end{lrbox}
\newsavebox{\SBoxSBThreeCount}
\begin{lrbox}{\SBoxSBThreeCount}\lstinline[language=Brbo]~$\clrthree{\cnt{\usebox{\SBoxSBThree}}}$~\end{lrbox}

\newsavebox{\SBoxSBOneStar}
\begin{lrbox}{\SBoxSBOneStar}\lstinline[language=Brbo]~|clr1:#sb1$\clrone{^\ubprime}$|~\end{lrbox}
\newsavebox{\SBoxSBTwoStar}
\begin{lrbox}{\SBoxSBTwoStar}\lstinline[language=Brbo]~|clr2:#sb2$\clrtwo{^\ubprime}$|~\end{lrbox}
\newsavebox{\SBoxSBThreeStar}
\begin{lrbox}{\SBoxSBThreeStar}\lstinline[language=Brbo]~|clr3:#sb3$\clrthree{^\ubprime}$|~\end{lrbox}

\newcommand{\resetleq}{\ensuremath{\color{dullteal}\leq}}



\newsavebox{\SBoxNumericalAbstractionFig}
\begin{lrbox}{\SBoxNumericalAbstractionFig}
\begin{lstlisting}[language=Java,alsolanguage=Brbo]
#sb := 0;
for ($\label{pt:na-before-i-init}$i := 0; i < #ts; i++) {
  p := 0;$\label{pt:na-before-p-init}$
  for (j := 0$\label{pt:na-before-inner}\lststopn$
       ; j < #tags; j++) {$\lststartn$
   {#sb$\,\le\,$(i$\cdot$#text+p)$\label{pt:na-inside-inner}$$\label{pt:na-complex-inv}\lststopn$$\rule{0pt}{3ex}$
         $+\,$((i$\cdot$#tags+j)$\cdot$ts#rep)$\rule{0pt}{2.5ex}$

         $+\,$(i$\cdot$#sep)}$\rule{0pt}{2.5ex}$

    l := *; r := *;$\rule{0pt}{3ex}$
     assume(p $\leq$ l $\leq$ r $\leq$ #text);$\lststartn$
    $\usebox{\SBoxSBUpdateOne}\label{pt:na-append-before-tag-before}$$\lststopn$
     assert(#sb $\le\bndsym$);$\lststartn$
    #rep := *;$\label{pt:na-append-before-tag-after}$$\lststopn$
     assume(0 $\leq$ #rep $\leq$ ts#rep);$\lststartn$
$\label{pt:na-append-rep-before}\lststopn$
    $\usebox{\SBoxSBUpdateTwo}$
     assert(#sb $\le\bndsym$);$\lststartn$
    p := r;
  }$\label{pt:exit-inner}$
  $\usebox{\SBoxSBUpdateThree}\label{pt:na-append-text-leftover-before}\lststopn$
   assert(#sb $\le\bndsym$);$\lststartn$
$\label{pt:na-append-text-leftover-after}\label{pt:na-append-separator-before}\lststopn$
  $\usebox{\SBoxSBUpdateFour}$
   assert(#sb $\le\bndsym$);$\lststartn$
}$\label{pt:na-append-separator-after}\label{pt:na-exit}$
\end{lstlisting}

\end{lrbox}

\newsavebox{\SBoxResetsFig}
\begin{lrbox}{\SBoxResetsFig}
\begin{lstlisting}[language=Java,alsolanguage=Brbo,numbers=left]

for (i := 0; i < #ts; i++) {$\label{pt:decomp-outer-loop}$
  p := 0;
  for (j := 0,$\label{pt:reset-sbone}\label{pt:decomp-inner-loop}\lststopn$
       $\usebox{\SBoxSBOneReset}$; j < #tags; j++) {$\lststartn$
  {$\usebox{\SBoxSBOneCount}$$\,=\,$i$\;\land\;\usebox{\SBoxSBOneStar}\le\,$#text$\;\land\;$#sb1$\,\le\,$p$\;\land$$\label{pt:decomp-complex-inv}\lststopn$$\rule{0pt}{3ex}$
   $\usebox{\SBoxSBTwoCount}$$\,=\,$i$\cdot$#tags+j-1$\;\land\;$$\rule{0pt}{2.5ex}$
   $\usebox{\SBoxSBTwoStar}\le\,$ts#rep$\;\land\;$#sb2$\,\le\,$ts#rep$\;\land\;$
   $\usebox{\SBoxSBThreeCount}$$\,=\,$i-1$\;\land\;$$\rule{0pt}{2.5ex}$
   $\usebox{\SBoxSBThreeStar}\le\,$#sep$\;\land\;$#sb3$\,\le\,$#sep}
    l := *; r := *;$\rule{0pt}{3ex}$
     assume(p $\leq$ l $\leq$ r $\leq$ #text);$\lststartn$
    $\usebox{\SBoxSBOneUpdateOne}$$\label{pt:ub-check-1}$$\lststopn$
     $\ubassertusecolor{\usebox{\SBoxSBOne},\usebox{\SBoxSBTwo},\usebox{\SBoxSBThree}}{\bndsym}$$\lststartn$
    #rep := *;$\lststopn$
     assume(0 $\leq$ #rep $\leq$ ts#rep);$\lststartn$
    $\usebox{\SBoxSBTwoReset}$$\label{pt:ub-check-2}$$\lststopn$
    $\usebox{\SBoxSBTwoUpdate}$
     $\ubassertusecolor{\usebox{\SBoxSBOne},\usebox{\SBoxSBTwo},\usebox{\SBoxSBThree}}{\bndsym}$$\lststartn$
    p := r;
  }
  $\usebox{\SBoxSBOneUpdateTwo}$$\label{pt:ub-check-3}$$\lststopn$
   $\ubassertusecolor{\usebox{\SBoxSBOne},\usebox{\SBoxSBTwo},\usebox{\SBoxSBThree}}{\bndsym}$$\lststartn$
  $\usebox{\SBoxSBThreeReset}$$\label{pt:ub-check-4}$$\lststopn$
  $\usebox{\SBoxSBThreeUpdate}$
   $\ubassertusecolor{\usebox{\SBoxSBOne},\usebox{\SBoxSBTwo},\usebox{\SBoxSBThree}}{\bndsym}$$\lststartn$
}$\label{pt:decomp-before-exit}$
\end{lstlisting}


\end{lrbox}

\newsavebox{\SBoxNumericalAbstractionPreCond}
\begin{lrbox}{\SBoxNumericalAbstractionPreCond}\lstinline[language=Brbo]~{0$\,\le\,$#text$\;\land\;$0$\,\le\,$#tags$\;\land\;$0$\,\le\,$#ts$\;\land\;$0$\,\le\,$ts#rep$\;\land\;$0$\,\le\,$#sep}~\end{lrbox}

\begin{figure}[tb]\centering
\begin{tabular}{r@{\quad}l}
\emph{global bound $\bndsym$}: & \usebox{\SBoxBoundRHS} \\
\emph{pre-condition}: & \usebox{\SBoxNumericalAbstractionPreCond}
\end{tabular}\par
\subfloat[A numerical abstraction of the \code{replaceTags} method from \figref{replaceTags-example}.]{\label{fig:inv-original}\usebox{\SBoxNumericalAbstractionFig}}
\hfill
\subfloat[A resource usage decomposition and amortized segmentation of (a).]{\label{fig:inv-simplified}\usebox{\SBoxResetsFig}}
\smallskip
\caption{Decomposing resource usage into amortized segments transforms the required supporting loop invariant at \ptref{na-complex-inv} needed to prove the global bound $\bndsym$ \emph{from polynomial to linear}.}
\label{fig:decomposition-example}
\end{figure}

Yet, our informal reasoning above did not require this level of complexity.
The key idea is that we can conceptually decompose the intermingled resource updates to \code{#sb} in any number of ways---and different decompositions select different amortizations.
In \figref{inv-simplified}, we illustrate a particular decomposition of updates to \code{#sb}.
We introduce three resource variables \code{|clr1:#sb1|}, \code{|clr2:#sb2|}, \code{|clr3:#sb3|} that correspond to the three parts of the informal argument above (\ie resource use for the non-tag text
at program points
\ref{pt:na-append-before-tag-before}
and \ref{pt:na-append-text-leftover-before}, the tag-replaced text
at \ptref{na-append-rep-before},
and the separator at \ptref{na-append-separator-before}, respectively).
Let us first ignore the $\resetkwusecolor$ and
$\ubassertkwusecolor$ commands (described further below), then we see that we are simply accumulating resource updates to \code{#sb} into separate variables or \emph{amortization groups}
such that \code{#sb}$\;=\;$\code{|clr1:#sb1|}$\,+\,$\code{|clr2:#sb2|}$\,+\,$\code{|clr3:#sb3|}.
But we can now bound \code{|clr1:#sb1|}, \code{|clr2:#sb2|}, and \code{|clr3:#sb3|} independently and have the sum of the bounds of these variables be a bound for the original resource variable \code{#sb}.

\newsavebox{\SBoxSBOneUpperBound}
\begin{lrbox}{\SBoxSBOneUpperBound}\lstinline[language=Brbo]~$\usebox{\SBoxSBOneCount} \cdot\usebox{\SBoxSBOneStar} +\,$#sb1~\end{lrbox}
\newsavebox{\SBoxSBTwoUpperBound}
\begin{lrbox}{\SBoxSBTwoUpperBound}\lstinline[language=Brbo]~$\usebox{\SBoxSBTwoCount} \cdot\usebox{\SBoxSBTwoStar} +\,$#sb2~\end{lrbox}
\newsavebox{\SBoxSBThreeUpperBound}
\begin{lrbox}{\SBoxSBThreeUpperBound}\lstinline[language=Brbo]~$\usebox{\SBoxSBThreeCount} \cdot\usebox{\SBoxSBThreeStar} +\,$#sb3~\end{lrbox}
\newsavebox{\SBoxSBUpperBound}
\begin{lrbox}{\SBoxSBUpperBound}\lstinline[language=Brbo]~#sb$\;\le\;(\usebox{\SBoxSBOneCount}\cdot\usebox{\SBoxSBOneStar} +\,$#sb1$) + (\usebox{\SBoxSBTwoCount}\cdot\usebox{\SBoxSBTwoStar} +\,$#sb2$) + (\usebox{\SBoxSBThreeCount}\cdot\usebox{\SBoxSBThreeStar} +\,$#sb3$)$~\end{lrbox}

However, precisely reasoning about the resource usage in \code{#sb1} still requires a polynomial loop invariant with the loop counters \code{i}, input \code{#text}, and internal variable \code{p}.
Following the observation from above, we want to amortize updates to \code{#sb1} across iterations of the inner loop but not between iterations of the outer loop.
That is, we want to amortize updates to \code{#sb1} in the sequence of resource uses within a single iteration of the outer loop and then apply worst-case reasoning to the resource bound amortized within this sequence.
The \emph{amortization reset} \usebox{\SBoxSBOneReset} after the initializer of the loop at \ptref{decomp-inner-loop} accomplishes this desired decoupling of the updates to \code{#sb1} between outer-loop iterations by ``resetting the amortization'' at each outer-loop iteration.
Conceptually, executions of the $\resetusecolor{\resvar}$ mark the boundaries of the \emph{amortization segments} of $\auxupdatekwusecolor$s of resource $\resvar$.

The result of this decomposition is the simpler invariant at \ptref{decomp-complex-inv} in the transformed program of \figref{inv-simplified}, which use some auxiliary summary variables like \usebox{\SBoxSBOneStar}
and \usebox{\SBoxSBOneCount}.
For every resource variable $\resvar$, we consider two summary variables $\ub\resvar$ and $\cnt\resvar$, corresponding, respectively, to the maximum of $\resvar$ in any segment and the number of ``$\resetkwusecolor$ted'' $r$ segments so far.
Concretely, the semantics of \usebox{\SBoxSBOneReset} is as follows:
\begin{inparaenum}[(1)]
\item increment the segment counter variable \usebox{\SBoxSBOneCount} by 1, thus tracking the number of amortization segments of \usebox{\SBoxSBOne} $\auxupdatekwusecolor$s;
\item bump up \usebox{\SBoxSBOneStar} if necessary (\ie set \usebox{\SBoxSBOneStar} to \lstinline[language=Brbo]~max($\usebox{\SBoxSBOneStar}$, #sb1)~), tracking the maximum \usebox{\SBoxSBOne} in any segment so far; and finally,
\item resets \code{#sb1} to 0 to start a new segment.
\end{inparaenum}
As we see at \ptref{na-complex-inv} in the original and transformed programs of \figref{decomposition-example}, we have decomposed the total non-tag text piece \lstinline[language=Brbo]~(i$\cdot$#text+p)~ into \usebox{\SBoxSBOneUpperBound} where
\lstinline[language=Brbo]~$\usebox{\SBoxSBOneCount}\,=\,$i~,
\lstinline[language=Brbo]~$\usebox{\SBoxSBOneStar}\le\,$#text~, and
\lstinline[language=Brbo]~#sb1$\,\le\,$p~.
Intuitively, $\usebox{\SBoxSBOneCount}\cdot\usebox{\SBoxSBOneStar}$ upper-bounds the cost of all \emph{past} iterations of the outer loop, and the cost of the \emph{current} iteration is precisely \code{#sb1}.
Thus, \usebox{\SBoxSBOneUpperBound} is globally and inductively an upper bound
for the total non-tag text piece of \code{#sb}.
The same decomposition applies to \usebox{\SBoxSBTwo} and \usebox{\SBoxSBThree} where note that
\lstinline[language=Brbo]~$\usebox{\SBoxSBThreeCount}\,=\,$i-1~,
counts past segments separated from the current segment so that \usebox{\SBoxSBThreeUpperBound} corresponds to
\lstinline[language=Brbo]~(i$\cdot$#sep)~ where both the past and current are summarized together.
Overall, combining the amortization groups and segments, we have the following global invariant between the original program and the transformed one:
\[
\usebox{\SBoxSBUpperBound}
\]

To verify a given bound in the transformed program, we simply check that this expression on the right in the above is bounded by the desired bound expression using any inferred invariants on \usebox{\SBoxSBOneCount}, \usebox{\SBoxSBOneStar}, \usebox{\SBoxSBOne}, etc.
This is realized by the \emph{upper-bound check} command at, for instance, \ptref{ub-check-1} in \autoref{fig:inv-simplified}:
\[
\ubassertusecolor{\usebox{\SBoxSBOne}, \usebox{\SBoxSBTwo}, \usebox{\SBoxSBThree}}{(\usebox{\SBoxBoundRHS})} \;.
\]
Here, $\ubassertusecolor{\resvarseq}{\expr}$ asserts that the sum of
amortization groups (internally decomposed into amortization segments) in the
set $\resvarseq$ is bounded from above by $e$. 


\subsection{Finding a Selective-Amortization Decomposition}
\label{subsec:overview-finding}

\Figref{inv-simplified} shows a decomposition of updates to \code{#sb} into groups (\ie \usebox{\SBoxSBOne}, \usebox{\SBoxSBTwo}, and \usebox{\SBoxSBThree}) and segments (\ie with $\resetkwusecolor$s) that realize a particular selective amortization.
We show that any decomposition into groups and segments is sound in \secref{decomposing}, but
here, we discuss how we find such a decomposition.



Intuitively, we want to use worst-case reasoning whenever possible, maximizing decoupling of updates and simplifying invariant inference.
But some updates should be considered together for amortization.
Thus, any algorithm to select a decomposition must attempt to resolve the tension between two conflicting goals: partitioning $\auxupdatekwusecolor$ updates in the program into more groups and smaller segments, but also allowing amortizing costs inside larger segments to avoid precision loss.
For example, it is important to use the same accumulation variable \usebox{\SBoxSBOne} for the two locations that contribute to the non-tag text (program points~\ref{pt:ub-check-1} and~\ref{pt:ub-check-3}) to amortize over both $\auxupdatekwusecolor$ sites.
In \secref{selecting-decomposition}, we characterize the potential imprecision caused by worst-case reasoning over segments with a notion of \emph{amortization segment non-interference}, which along with some basic restrictions motivates the approach we describe here.

{
\newcommand{\one}{\ref{pt:na-append-before-tag-before}}
\newcommand{\two}{\ref{pt:na-append-text-leftover-before}}

\begin{figure}[b]
  \centering
\scalebox{0.79}{
  \begin{tikzpicture}[every text node part/.style={align=center}]
      \node[state,shape=ellipse,initial
        left,fill=safecellcolor,
        label=below:${\one: {\code{()}}}$
        \\${\two: {\code{()}}}$
        ,draw] (S0) at (0,0) {$\ref{pt:na-before-i-init}$};
      \node[state,shape=ellipse,fill=safecellcolor,
        label=below:${\one: {\code{()}}}$
          \\${\two: {\code{()}}}$
        ,draw] (S1) [right=2cm of S0] {$\ref{pt:na-before-i-init}^\ast$};
      \node[state,shape=ellipse] (S2)
           [right=2.5cm of S1] {$\ref{pt:na-exit}^\ast$};
      \node[state,shape=ellipse,fill=safecellcolor,
        label=below:${\one: {\code{()}}}$
          \\${\two: {\code{()}}}$
        ,draw] (S3) [below left=4cm of S1] {$\ref{pt:na-before-p-init}$};
      \node[state,shape=ellipse,fill=safecellcolor,
        label=below:${\one: {\code{(p: $0$)}}}$
          \\${\two: {\code{(p: $0$)}}}$
        ,draw] (S4) [right=1.6cm of S3] {$\ref{pt:na-before-inner}$};
      \node[state,shape=ellipse,fill=goodcellcolor,
        label=below:${\one: {\code{(p: $0$)}}}$
          \\${\two: {\code{(p: $0$)}}}$
        ,draw] (S5) [right=1.6cm of S4] {$\ref{pt:na-before-inner}^\ast$};
      \node[state,shape=ellipse] (S6) [right=4cm of S1,
        label=below:${\one: {\code{()}}}$
          \\${\two: {\code{(p: $\top$)}}}$
        ,draw] {$\ref{pt:na-append-text-leftover-before}$};
      \node[state,shape=ellipse] (S7) [right=2.3cm of S5,
        label=below:${\one: {\code{(p: $\top$)}}}$
          \\${\two: {\code{(p: $\top$)}}}$
        ,draw] {$\ref{pt:na-inside-inner}$};
      \node[state,shape=ellipse] (S8) [right=4cm of S6,
        label=below:${\one: {\code{()}}}$
          \\${\two: {\code{()}}}$
        ,draw] {$\ref{pt:na-append-text-leftover-after}$};
      \node[state,shape=ellipse,draw=none] (S12) [right=1cm of S8] {$\dots$};
      \node[state,shape=ellipse] (S14) [right= 1cm of S12,
        label=below:${\one: {\code{()}}}$
          \\${\two: {\code{()}}}$
        ,draw] {$\ref{pt:na-append-separator-after}$};
      \node[state,shape=ellipse] (S9) [right=3cm of S7,
        label=below:${\one: {\code{(l: $\top$, p: $\top$)}}}$
          \\${\two: {\code{(p: $\top$)}}}$
        ,draw] {$\ref{pt:na-append-before-tag-before}$};
      \node[state,shape=ellipse] (S11) [right=3.4cm of S9,
        label=below:${\one: {\code{(r: $\top$)}}}$
        \\${\two: {\code{(r: $\top$)}}}$
        ,draw] {$\ref{pt:na-append-before-tag-after}$};
      \node[state,shape=ellipse,draw=none] (S13) [right=1cm of S11] {$\dots$};
      \node[state,shape=ellipse] (S15) [right=1cm of S13,
        label=below:${\one: {\code{(p: $\top$)}}}$
          \\${\two: {\code{(p: $\top$)}}}$
        ,draw] {$\ref{pt:exit-inner}$};

      \path[->]
      (S0) edge node {$\code{i:=0}$} (S1)
      (S1) edge node[above] {$\code{i}{\geq}\code{\#ts}$} (S2)
      (S1) edge node[right] {$\code{i}{<}\code{\#ts}$} (S3)
      (S3) edge node[below] {$\code{p:=0}$} (S4)
      (S4) edge node[below] {$\code{j:=0}$} (S5)
      (S5) edge node[left] {$\code{j}{\geq}\code{\#tags}$} (S6)
      (S5) edge node[below] {$\code{j}{<}\code{\#tags}$} (S7)
      (S6) edge node[above] {$\auxupdate{\usebox{\SBoxSBOne}}{\usebox{\SBoxTextMinusPOp}}$} (S8)
      (S8) edge node {} (S12)
      (S12) edge node {} (S14)
      (S7) edge node[below] {$\code{p}{\le}\code{l}{\le}\code{r}{\le}\code{\#text}$} node[above] {$\code{l:=*; r:=*;}$} (S9)
      (S9) edge node[below] {$\auxupdate{\usebox{\SBoxSBOne}}{\usebox{\SBoxLMinusPOp}}$} (S11)
      (S11) edge node {} (S13)
      (S13) edge node {} (S15)
      (S14) edge[bend right=25] node {} (S1)
      (S15) edge[bend right=20] node {} (S5)
    
      ;
   \end{tikzpicture}
  }\smallskip
\caption{%
Inserting a \usebox{\SBoxSBOneReset} to select a segmentation for amortization group \usebox{\SBoxSBOne}.
We show the program from \figref{inv-simplified} here as a control-flow graph.
}
\label{fig:cfg}
\end{figure}
}

In \figref{cfg}, we show the control-flow graph of the resource-decomposed program in \figref{inv-simplified} without the inserted $\resetkwusecolor$s.
Node labels correspond to program points there, except for labels
$\ref{pt:decomp-outer-loop}^\ast$, $\ref{pt:decomp-inner-loop}^\ast$, $\ref{pt:decomp-before-exit}^\ast$ that correspond to unlabeled program points in the initialization of the \code{for}-loops and the procedure exit.
Edges are labeled by a single or a sequence of commands (where we omit keyword \code{assume} for brevity in the figure).
Some nodes and edges are elided as $\ldots$ that are not relevant for this discussion.
Ignore node colors and the labels below the nodes for now.

Let us consider the class of \emph{syntactic} selective-amortization transformations where we can rewrite resource use commands $\auxupdateusecolor{\resvar}{\expr}$ to place $\auxupdatekwusecolor$s into separate amortization groups, and we can insert a $\resetusecolor{\resvar'}$ at a single program location to partition $\auxupdatekwusecolor$s into amortization segments for each group $\resvar'$.
But otherwise, we make no other program transformation.
We then use the notion of segment non-interference to select a group and segment decomposition under this syntactic restriction.

Now, the intuition behind amortization segment non-interference is that two segments for a resource $\resvar$ are non-interfering if under the same ``low inputs,'' the resource usage of $\resvar$ is the same in both segments.
In \figref{cfg}, the labels below the nodes show such low inputs to a particular $\auxupdatekwusecolor$ site from a particular program point.
For example, under node~\ref{pt:decomp-inner-loop}, we show \code{p} as a low input for both the $\auxupdatekwusecolor$ sites at program points~\ref{pt:ub-check-1} and~\ref{pt:ub-check-3} (ignore the \code{:$0$}s for the moment).

So, an additional parameter in our search space is a partitioning of variables into ``low'' and ``high'' ones (which we note are not distinguished based on security relevance in the standard use of the non-interference term~\cite{DBLP:conf/popl/AbadiBHR99} but rather on relevance for amortized reasoning).
We further fix the low variables in any segmentation we might use to be the internal variables on which the $\auxupdatekwusecolor$s data-depend. This is based on the intuition that $\auxupdatekwusecolor$s that share computation over internal variables are related for amortization.
Because the $\auxupdatekwusecolor$s for \usebox{\SBoxSBOne} at program points~\ref{pt:ub-check-1} and~\ref{pt:ub-check-3} share \code{p} as an input at, for example, node~\ref{pt:decomp-inner-loop}, we place these $\auxupdatekwusecolor$ sites in the same group. Then, otherwise the other $\auxupdatekwusecolor$ sites at program points~\ref{pt:ub-check-2} and~\ref{pt:ub-check-4} are placed in other groups (namely, \usebox{\SBoxSBTwo} and \usebox{\SBoxSBThree}, respectively).
The set of variables on which $\auxupdatekwusecolor$ sites data-depend can be computed by a standard program slicing~\cite{DBLP:journals/tse/Weiser84}.

Finally, we insert a single $\resetkwusecolor$ for each group to define amortization segments.
So that all $\auxupdateusecolor{\resvar}{\expr}$ commands for a group $\resvar$ are always after some $\resetusecolor{\resvar}$, we consider program locations that control-dominate all $\auxupdatekwusecolor$ sites for $\resvar$.
In \figref{cfg}, any of the colored nodes control-dominate the two $\auxupdatekwusecolor$ sites for \usebox{\SBoxSBOne}.
To make the amortization segments as small as possible (while minimizing precision loss), we select the most immediate dominator where the low variables can be proven constant (\ie the low inputs to the segments will always the same value).
Node $\ref{pt:decomp-inner-loop}^\ast$ (colored green) is this dominator for the two $\auxupdatekwusecolor$ sites for \usebox{\SBoxSBOne} because \code{p} is always $0$ (shown as \code{p:$0$}) and where we insert $\resetusecolor{\usebox{\SBoxSBOne}}$.
We can derive this constancy property with any numerical abstract domain (here, we show $\top$ for non-constant values from a standard constant propagation analysis for presentation), and we can pessimistically assume other variables to be low and also try to prove constancy for them to potentially recover some additional precision in segmentation.

Note that the analyses being applied here are classical ones. What is interesting here is not the analyses per se but their application to selecting amortization groups and segments
to realize selectively-amortized resource bounding.

\section{Decomposing Resource Usage}
\label{sec:decomposing}

Our technique considers a resource-usage tracking program and splits a single resource variable into an arbitrary number of \emph{resource decompositions}.
By design, resource-usage tracking updates are generic in allowing updates with any integer-valued expression, enabling modeling non-monotonic resources like list additions and removals or memory allocation and deallocation. 
In this section, we define a core imperative language for resource-usage tracking, formalize selective-amortized analysis as a program transformation that inserts \emph{amortization resets} into decomposed resource-usage tracking variables (\secref{resetting}), and show that any transformation is sound with respect to bound checks on resource usage (\secref{resetting-soundness}).
While we focus on upper-bound checks, we will see that the approach can be easily adapted for lower-bound assertions.

\newsavebox{\SBoxCoreLangFig}
\begin{lrbox}{\SBoxCoreLangFig}\small
\begin{mathpar}
\text{values $\val \bnfdef \valn \bnfalt \valb \bnfalt \cdots$}

\text{booleans $\valb \bnfdef \kwtrue \bnfalt \kwfalse$}

\text{expressions $\expr \bnfdef \var \bnfalt \val \bnfalt \cdots$}

\begin{grammar}[r]
  commands & \cmd & \bnfdef &
  \cmdskip
  \bnfalt \assign{\var}{\expr}
  \bnfalt \assume{\expr}
  \emcolor\bnfalt \auxupdate{\resvar}{\expr}
  \bnfalt \ubassert{\resvarseq}{\expr}
  \bnfalt \reset{\resvar} 
  \\
  programs & \prog & \bnfdef & \emp \bnfalt \transcons{\prog}{\mktrans{\loc}{\cmd}{\loc'}}
\end{grammar} \\

\text{variables $\var$}

\text{resources $\resvar$}

\text{locations $\loc$}

\begin{grammar}[r]
  stores & \pstore & \bnfdef &
  \emp
  \bnfalt \mapext{\pstore}{\var}{\val}
  \emcolor\bnfalt \mapext{\pstore}{\resvar}{\valn}
  \bnfalt \mapext{\pstore}{\ub{\resvar}}{\valn}
  \bnfalt \mapext{\pstore}{\cnt{\resvar}}{\valn}
\end{grammar} \\

\fbox{\evalexpr{\pstore}{\expr}{\val}

\evalcmd{\pstore}{\cmd}{\pstore'}} \\

\infer[E-Use]{
  \evalexpr{\pstore}{\expr}{\valn}
}{
  \evalcmd{\pstore}{ \auxupdate{\resvar}{\expr} }{ \mapext{\pstore}{\resvar}{\pstore(\resvar) + \valn} }
}

\infer[E-UBCheck]{
  \evalexpr{\pstore}{\expr}{\valn}
  \\
  \left(\Sum_{\resvar \in \resvarseq}
  \pstore(\cnt{\resvar}) \cdot \pstore(\ub{\resvar}) + \pstore(\resvar)\right)
  \leq \valn
}{
  \evalcmd{\pstore}{ \ubassert{\resvarseq}{\expr} }{ \pstore }
}

\infer[E-Reset]{
  \pstore' =
  \mapext{
    \mapext{
      \mapext{
        \pstore}
             {\cnt{\resvar}}{ \pstore(\cnt{\resvar}) + 1}
    }
           {
             \ub{\resvar}}{ \max(\pstore(\ub{\resvar}), \pstore(\resvar))
           }
  }{\resvar}{0}
}{
  \evalcmd{\pstore}{ \reset{\resvar} }{ \pstore' }
}
\end{mathpar}
\end{lrbox}
\begin{figure}[tb]\centering
\usebox{\SBoxCoreLangFig}
\smallskip
\caption{A core imperative language for resource-usage analysis. Resources
$\resvar$ are modeled as integer-valued variables that may increase or
decrease (via a $\auxupdatekw$ command) and bound-checked (via an
$\ubassertkw$ assertion command). Selective amortization is realized through
\emph{resource} $\resetkw$\emph{s}.}
\vspace{-1em}
\label{fig:corelang}
\end{figure}

In \figref{corelang}, we give the core resource-usage tracking language.
We consider an unspecified expression language $\expr$, aside from including program variables $\var$ and its value forms $\val$ having integers $\valn$ and booleans $\valb$.
The command forms include standard imperative ones like the no-op unit $\cmdskip$, assignment $\assign{\var}{\expr}$, and guard condition $\assume{\expr}$.
The remaining \textemcolor{highlighted} command forms work with resources $\resvar$.
In particular, $\auxupdate{\resvar}{\expr}$ models a resource use where the usage of $\resvar$ is incremented by the value of $\expr$, and $\ubassert{\resvarseq}{\expr}$ is an upper-bound assertion checking that the sum of the resources $\resvarseq$ is upper-bounded by the value of $\expr$.
We abuse notation slightly by writing $\resvarseq$ both for a sequence $\resvar_1 \ldots \resvar_n$ or a set $\Set{\resvar_1, \ldots, \resvar_n}$ of resources.
Selective amortization is realized through resetting resources with the $\reset{\resvar}$ command that we detail further below.
Note that program expressions $\expr$ do not contain resources variables $\resvar$.
Finally, programs $\prog$ are given as control-flow graphs with edges $\mktrans{\loc}{\cmd}{\loc'}$ labeled by commands~$\cmd$ between locations~$\loc$.

The states $\pstate$ of a program are pairs $\locstore{\loc}{\pstore}$ of locations $\loc$ and stores $\pstore$.
Stores are finite maps, mapping program variables to values $\ext{\var}{\val}$, as well as tracking resources in the remaining \textemcolor{highlighted} forms.
A resource $\resvar$ is a integer-valued variable $\ext{\resvar}{\valn}$.
For any resource $\resvar$, we consider two auxiliary resource-usage summary
variables $\ub{\resvar}$ and $\cnt{\resvar}$ used in resource resetting to be
described later. 

A judgment form for evaluating expressions $\evalexpr{\pstore}{\expr}{\val}$
stands for ``In store $\pstore$, expression $\expr$ evaluates to value $\val$.''
Similarly, a judgment form $\evalexpr{\pstore}{\cmd}{\pstore'}$ stands for ``In store $\pstore$, command $\cmd$ updates the store to $\pstore'$.''
In \figref{corelang}, we elide the standard rules for $\cmdskip$, assignment $\assign{\var}{\expr}$, and guard condition $\assume{\expr}$ and focus on the resource-manipulating commands.

The \TirName{E-Use} rule captures that the $\auxupdate{\resvar}{\expr}$ command says to increment $\resvar$ by the value of $\expr$.
Note that we write $\pstore(\resvar)$ for looking up the mapping of $\resvar$ in store $\pstore$ and assume that any unmapped $\resvar$ maps to $0$. That is, we consider all resources $\resvar$ initialized to $0$.
The \TirName{E-UBCheck} describes an upper-bound check $\ubassert{\resvarseq}{\expr}$ on a set of resources $\resvarseq$.
Let us first consider a single resource $\resvar$ and assume that the auxiliary variable $\ub{\resvar}$ is 0 in the store.
Then, the rule simply checks that $\resvar$ is upper-bounded by the value of $\expr$ (\ie like $\assert{\resvar \leq \expr}$).
In the next subsection, we come back to the more general form of the upper-bound check shown in \TirName{E-UBCheck}, which captures the essence of selectively-amortized resource bounding through an interaction with resource decomposition and amortization resets.

\subsection{Selective Amortization By Decomposition}
\label{sec:resetting}

Recall from \secref{overview} that the essence of selectively-amortized resource bounding is we want to selectively choose the sequence of resource uses $\auxupdate{\resvar}{\expr}$ over which we apply amortized reasoning.
To do this, we have two intertwined tools: resource decomposition $\decompto{\resvar}{\resvarseq}$ into \emph{amortization groups} and amortization resets $\reset{\resvar}$ into \emph{amortization segments}.

\newsavebox{\SBoxDecompFig}
\begin{lrbox}{\SBoxDecompFig}\small
\begin{mathpar}
\begin{grammar}[l]
  decompositions & \decomp & \bnfdef & \emp \bnfalt \decompcons{\decomp}{\decompto{\resvar}{\resvarseq}} \\
\end{grammar} \\

\fbox{%
\hypdecomp{\cmd}{\cmd'}} \\

\infer[D-Use]{
  \resvar' \in \resvarseq
}{
  \hypdecomp[\decompcons{\decomp}{\decompto{\resvar}{\resvarseq}}]{ \auxupdate{\resvar}{\expr} }{ \auxupdate{\resvar'}{\expr} }
}

\infer[D-UBCheck]{
}{
  \hypdecomp[\decompcons{\decomp}{\decompto{\resvar}{\resvarseq}}]{ \ubassert{\resvar}{\expr} }{ \ubassert{\resvarseq}{\expr} }
}

\infer[D-Reset]{
}{
   \hypdecomp{ \cmdskip }{ \reset{\resvar} }
}

\infer[D-Command]{
  \cmd \in \Set{ \cmdskip, \assign{\var}{\expr}, \assume{\expr}}
}{
  \hypdecomp{ \cmd }{ \cmd }
}
\end{mathpar}
\end{lrbox}
\begin{figure}[tb]\centering
\usebox{\SBoxDecompFig}
\smallskip
\caption{Decomposing resource usage for selective-amortization analysis is
described with a transformation that rewrites commands with
a resource decomposition $\decomp$.
Decompositions $\decomp$ define the amortization groups, while inserted
$\resetkw$s determine the amortization segments.}
%
\label{fig:decomp}
\end{figure}

A resource decomposition $\decomp \bnfdef \emp \bnfalt \decompcons{\decomp}{\decompto{\resvar}{\resvarseq}}$ is a mapping from a resource $\resvar$ into a set of decomposed resource-usage tracking variables $\resvarseq$.
The transformation takes $\auxupdate{\resvar}{\expr}$ and rewrites them to use
$\auxupdate{\resvar'}{\expr}$ for some $\resvar' \in \resvarseq$, thus
decomposing all uses of $\resvar$ into separate amortization groups given by
$\resvarseq$. 
In \figref{decomp}, the judgment form $\hypdecomp{\cmd}{\cmd'}$ says, ``Under
resource decomposition $\decomp$, command $\cmd$ can be resource-decomposed to
command $\cmd'$,'' stating valid decomposition transformations. 
The \TirName{D-Use} rule states exactly this transformation for
$\auxupdate{\resvar}{\expr}$ commands. 

Then, within separate amortization groups, resets $\reset{\resvar}$ define the
\emph{segments} of execution over which to amortize resource uses while applying
worst-case reasoning around them.
To see this, consider the \TirName{E-Reset} rule in \figref{corelang} where we
can see $\reset{\resvar}$ as corresponding to the following assignments (abusing notation slightly with assignments and expressions using
resource variables): 
\[
\quad\assign{\cnt{\resvar}}{\cnt{\resvar} + 1};
\quad\assign{\ub{\resvar}}{\max(\ub{\resvar}, \resvar)};
\quad\assign{\resvar}{0};
\]
That is, the $\reset{\resvar}$ command
increments the number of amortization segments for $\resvar$ seen so far in $\cnt{\resvar}$,
saves the maximum value of $\resvar$ in any segment so far in $\ub{\resvar}$, and
resets $\resvar$ to 0 ending the last amortization segment and starting the next one.
So the $\ub{\resvar}$ resource-usage summary captures the worst-case resource use of $\resvar$ over all segments, while the $\cnt{\resvar}$ summary saves the number of such amortization segments.

These summaries then enables amortized reasoning within segments and worst-case reasoning around them.
To see this, let us consider a one-to-one resource decomposition $\decompto{\orig{\resvar}}{\resvar}$.
Without loss of generality, we assume the original program using $\orig{\resvar}$ does not have any $\resetkw$s (but the transformed program with $\resvar$ may). Furthermore, we assume all amortization segments are paths of the form $\traceconscmd{\traceconsstate{\pstore}{\reset{\resvar}} \cdots}{\pstore'}{\reset{\resvar}}$ with no other $\reset{\resvar}$ in the middle and that there are no resource uses $\auxupdate{\resvar}{\expr}$ before an initial $\reset{\resvar}$ (\ie all executions of $\auxupdate{\resvar}{\expr}$ are either in a segment bracketed by two $\reset{\resvar}$s or after the last $\reset{\resvar}$).
Then the following selective-amortization assertion between $\orig\resvar$ and $\resvar$ holds globally (in all reachable stores):
\[
  \orig{\resvar} \quad\leq\quad \cnt{\resvar} \cdot \ub{\resvar} + \resvar
\]
Intuitively, up to the last $\reset{\resvar}$, there have been $\cnt{\resvar}$ amortization segments and the worst-case use of $\resvar$ on all prior segments is $\ub{\resvar}$, so $\cnt{\resvar} \cdot \ub{\resvar}$ is an upper bound on the resource use up to the the last $\reset{\resvar}$---thereby using worst-case reasoning on amortized segments.
Then we just add $\resvar$ because the remaining uses $\auxupdate{\resvar}{\expr}$ since the last reset have accumulated in $\resvar$.
Note that we thus consider all upper-bound summaries $\ub{\resvar}$ initialized to 0 and all segment-counter summaries $\cnt{\resvar}$ initialized to -1.

Coming back to the \TirName{E-UBCheck} rule describing the upper-bound check $\ubassert{\resvar}{\expr}$ in \figref{corelang} (for a single resource $\resvar$), the assertion checks the bound $\expr$ on exactly this amortized segments expression (\ie like $\assert{\cnt{\resvar} \cdot \ub{\resvar} + \resvar \leq \expr$}).
Then, with respect to amortization groups, a resource decomposition $\decompto{\resvar}{\resvarseq}$ says that resource uses to $\resvar$ are distributed over uses to $\resvarseq$, so we simply sum over the amortization groups $\resvarseq$ (\ie like $\assert{\left(\sum_{\resvar \in \resvarseq} \cnt{\resvar} \cdot \ub{\resvar} + \resvar \right) \leq \expr}$).

Thus, the transformation from an upper-bound check $\ubassert{\resvar}{\expr}$ on a resource $\resvar$ with decomposition $\decompto{\resvar}{\resvarseq}$ yields $\ubassert{\resvarseq}{\expr}$ as stated in rule \TirName{D-UBCheck} from \figref{decomp}.
As alluded to above, it is sound to insert $\resetkw$s arbitrarily into the transformed program corresponding to different amortization segments, which we state with rule \TirName{D-Reset}.
Note that we consider programs $\prog$ equivalent up to insertions of $\cmdskip$ commands, so we can insert them into the original program as needed.
The remaining non-resource manipulating commands are simply retained as-is with rule \TirName{D-Command}.
For simplicity in presentation, we assume the original program does not have $\resetkw$s and
has only single-resource upper-bound checks $\ubassert{\resvar}{\expr}$.
Overall, any choice of a resource decomposition $\decomp$ is sound corresponding to different amortization groups. Again for simplicity, we assume all resources $\resvar$ in the original program have a mapping in $\decomp$ (e.g., at least have $\decompto{\resvar}{\resvar}$ for no decomposition).
We consider soundness in more detail further below.

\subsection{Soundness of Group and Segment Decomposition}
\label{sec:resetting-soundness}

To consider the soundness of the resource decomposition transformation
$\hypdecomp{\cmd}{\cmd'}$, we define program executions or paths
$\ptrace$. 
In \figref{paths}, we define paths $\ptrace$ in a slightly non-standard way:
they are sequences created by appending a state
$\traceconsstate{\ptrace}{\pstate}$ or appending a store-command pair
$\traceconscmd{\ptrace}{\pstore}{\cmd}$ and are well-formed if they consist of
sequences corresponding to the stores from valid executions of the commands (as
captured by the $\ptraceok{\ptrace}$ judgment). 
Intentionally, we define paths mostly independent from programs, stripping out
locations $\ell$ except for the last state $\locstore{\ell}{\pstore}$. 
In most cases, we do not care about the program from which paths may come from. 
For example, the path well-formedness judgment $\ptraceok{\ptrace}$ ignores
program locations and simply checks that the triples of store $\pstore$, command
$\cmd$, and store $\pstore'$ are valid executions
$\evalcmd{\pstore}{\cmd}{\pstore'}$ (rule \TirName{Ok-Step}). 
Unless otherwise stated, we assume all paths $\ptrace$ are well formed (\ie
$\ptraceok{\ptrace}$ holds for any path $\ptrace$). 

\newsavebox{\SBoxSoundnessFig}
\begin{lrbox}{\SBoxSoundnessFig}\small
\begin{mathpar}
\text{states $\pstate \bnfdef \locstore{\loc}{\pstore}$}

\text{paths $\ptrace \in \ptraceset \bnfdef
  \emp
  \bnfalt \traceconsstate{\ptrace}{\pstate}
  \bnfalt \traceconscmd{\ptrace}{\pstore}{\cmd}$} \\

\fbox{\ptraceok{\ptrace}

\stepprog{\pstate}{\pstate'}

$\denote{\prog} \pstate = \ptraceset$} \\

\infer[Ok-Init]{
}{
  \ptraceok{\pstate}
}

\infer[Ok-Step]{
  \ptraceok{ \traceconsstate{\ptrace}{\locstore{\loc}{\pstore}} } \\
  \evalcmd{\pstore}{\cmd}{\pstore'}
}{
  \ptraceok{
    \traceconsstate{\traceconscmd{\ptrace}{\pstore}{\cmd}}{\locstore{\loc'}{\pstore'}} 
  }
}

\infer[Step]{
  \mktrans{\loc}{\cmd}{\loc'} \in \prog \\
  \evalcmd{\pstore}{ \cmd }{ \pstore' }
}{
  \stepprog{ \locstore{\loc}{\pstore} }{ \locstore{\loc'}{\pstore'} }
}

\denote{\prog} \pstate \defeq
\lfp \lambda \ptraceset. \Set{ \pstate }
\union \Union_{\traceconsstate{\ptrace}{\locstore{\loc}{\pstore}} \in \ptraceset}
\SetST{ \traceconsstate{\traceconscmd{\ptrace}{\pstore}{\cmd}}{\pstate'} }{ \stepprog{\locstore{\loc}{\pstore}}{\pstate'} } \\

\fbox{%
 \hypdecomp{\ptrace}{\ptrace'}} \\



\infer[D-AppendCommand]{
  \hypdecomp{
    \traceconsstate{\ptrace}{\locstore{\ell}{\pstore}}
  }{
    \traceconsstate{\ptrace'}{\locstore{\ell'}{\pstore'}}
  }
  \quad
  \hypdecomp{\cmd}{\cmd'}
}{
  \hypdecomp{
    \traceconscmd{\ptrace}{\pstore}{\cmd}
   }{
    \traceconscmd{\ptrace'}{\pstore'}{\cmd'}
   }
}
\quad
\infer[D-Step]{
  \hypdecomp{\ptrace}{\ptrace'}
  \quad
  \ptraceok{\traceconsstate{\ptrace'}{\pstate'}}
  \quad
  \samevarstore{\pstate}{\pstate'}
}{
  \hypdecomp{
    \traceconsstate{\ptrace}{\pstate}
   }{
    \traceconsstate{\ptrace'}{\pstate'}
   }
}
\quad
\infer[D-Init]{
}{
  \hypdecomp{\pstate}{\pstate}
}

\fbox{%
\samevarstore{\pstore}{\pstore'}

\samevarstore{\pstate}{\pstate'}} \\

\begin{tabular}{r@{\;\;}c@{\;\;}l}
$\samevarstore{\orig\pstore}{\pstore}$ & iff &
  $\orig\pstore(\var) = \pstore(\var)$ for all $\var \in \Vars(\orig\pstore) = \Vars(\pstore)$ and \\[1ex]
  & &
  $\orig\pstore(\orig\resvar)
  \leq
  \Sum_{\resvar \in \decomp(\orig\resvar)}
    \pstore(\cnt{\resvar}) \cdot \pstore(\ub{\resvar}) + \pstore(\resvar)$
  for all $\orig\resvar \in \Domain(\orig\pstore)$
\end{tabular}

\text{$\samevarstore{\locstore{\loc}{\pstore}}{\locstore{\loc'}{\pstore'}} \;\;\text{iff}\;\; \samevarstore{\pstore}{\pstore'}$}

\Vars(\pstore) \defeq \SetST{ \var }{ \var \in \Domain(\pstore) } \\

\fbox{%
\hypdecomp{\prog}{\prog'}
} \\

\infer[D-Transition]{
  \hypdecomp{\cmd}{\cmd'}
}{
  \hypdecomp{
    \transcons{\prog}{ \mktrans{\loc}{\cmd}{\loc'} }
  }{
    \transcons{\prog'}{ \mktrans{\loc}{\cmd'}{\loc'} }
  }
}

\infer[D-EmptyProgram]{
}{
  \hypdecomp{ \emp }{ \emp }
}
\end{mathpar}
\end{lrbox}
\begin{figure}[tb]
\usebox{\SBoxSoundnessFig}
\smallskip
\caption{%
A semantic decomposition is captured with a path transformation $\hypdecomp{\ptrace}{\ptrace'}$ where paths $\ptrace$ are sequences of command executions.
The path transformation says we can rewrite according to the command transformation until reaching the same initial state.
That is, choosing amortization groups with any decomposition $\decomp$ and amortization segments with any insertions of $\resetkw$s are sound.
A syntactic decomposition is simply a lifting of the command transformation to programs $\hypdecomp{\prog}{\prog'}$ on the same control-flow structure.
}
\label{fig:paths}
\end{figure}

The only reason paths mention locations is to define the path semantics $\denote{\prog} \pstate$ of a program $\prog$ with initial state $\sigma$.
The path semantics $\denote{\prog} \pstate$ is given as:
\begin{inparaenum}[(1)]
  \item the judgment form $\stepprog{\pstate}{\pstate'}$ defines a transition relation saying, ``On program $\prog$, state $\pstate$ steps to state $\pstate'$,'' and
  \item the path semantics $\denote{\prog} \pstate$ collects all finite (but unbounded) prefixes of the transition system from the initial state $\pstate$.
\end{inparaenum}

The judgment form $\hypdecomp{\ptrace}{\ptrace'}$ states a selectively-amortized resource bounding on a path $\ptrace'$ from an original path $\ptrace$.
Divorcing paths from programs emphasizes that semantically, we can choose any amortization grouping with a choice of the resource decomposition $\decomp$ and select any amortization segmentation by inserting $\resetkw$s anywhere along the original path $\ptrace$.
The \TirName{D-AppendCommand} rules says that a command along the original path can be rewritten according to the command transformation $\hypdecomp{\cmd}{\cmd'}$.
Note that like with programs, we consider paths $\ptrace$ equivalent up to insertions of $\cmdskip$ commands, so we can insert them into the original path as needed. 

To talk about resource-decomposed stores along paths, we define
$\samevarstore{\pstore}{\pstore'}$ to be stores that are equal on program
variables $\Vars(\pstore)$ (excluding resource variables $\resvar$) and whose resource-usage tracking variables satisfy the
selectively-amortized assertion from \secref{resetting} (see \figref{paths} for
a detailed definition). 
Then, the \TirName{D-Step} rule says that the execution of the last command in
$\ptrace'$ must result in a state $\pstate'$ consistent with the semantics of
commands ($\ptraceok{\traceconsstate{\ptrace'}{\pstate'}}$) and with selective
amortization ($\samevarstore{\pstate}{\pstate'}$). 
Finally, the \TirName{D-Init} rule simply says that resource-decomposed paths should start with the same initial state. 

We can then consider a more restricted, syntactic class of selectively-amortized resource-bounding transformations by simply transforming the commands of a program $\prog$ (\ie the judgment form $\hypdecomp{\prog}{\prog'}$ in \figref{paths}).
To achieve more semantic selective amortizations, one could, of course, first apply richer semantics-preserving program transformations to the original program (than inserting $\cmdskip$s) before applying the resource-decomposition transformation.

We can now state the following soundness result.
\begin{theorem}[Soundness of Selectively-Amortized Resource Bounding\label{thm:soundness}]
\begin{enumerate}\itemsep 1ex
\item\label{soundness-cmds} If $\hypdecomp{\orig\cmd}{\cmd}$,
      $\evalcmd{\pstore}{\cmd}{\pstore'}$, and
      $\samevarstore{\orig\pstore}{\pstore}$,
      then $\evalcmd{\orig\pstore}{\orig\cmd}{\orig{\pstore'}}$
      with $\samevarstore{\orig{\pstore'}}{\pstore'}$.
\item\label{soundness-paths} If $\hypdecomp{\orig\ptrace}{\ptrace}$ and $\ptraceok{\ptrace}$,
      then $\ptraceok{\orig\ptrace}$.
\item\label{soundness-programs} If $\hypdecomp{\orig\prog}{\prog}$
      and $\ptrace \in \denote{\prog} \pstate$,
      then there is a $\orig\ptrace \in \denote{\orig\prog} \pstate$
      s.t. $\hypdecomp{\orig\ptrace}{\ptrace}$.
\end{enumerate}
\end{theorem}
The key lemma (part \ref{soundness-cmds}) states a preservation property
that any command decomposition preserves the selectively-amortized
resource-bounding invariant $\samevarstorerel$ (see \refappendix{sec:proofs} for details). 

\paragraph{Verifying Bounds with Selective Amortization.}
Bound verification by selective amortization follows directly from the soundness theorem given above.
In particular, given a particular resource composition $\decomp$ and a transformed program $\prog$ from the original program $\orig\prog$ such that $\hypdecomp{\orig\prog}{\prog}$, simply apply any off-the-shelf numerical verification or invariant generator to $\prog$ to try to prove translated upper-bound assertions $\ubassert{\resvarseq}{\expr}$ in $\prog$.

In \secref{selecting-decomposition}, we describe an approach for selecting a resource decomposition and inserting amortization resets.
However, we note that our key contribution described here is generically defining the space of selective amortizations.

\paragraph{Lower Bounds.}

While we focused on upper-bound checks in this section, we see that the approach can be adapted to lower-bound assertions in a straightforward manner by introducing a lower-bound resource-usage summary variable, say $\lb{\resvar}$.
This lower-bound summary is analogously updated on $\reset{\resvar}$ with the minimum resource-usage so far (\ie like $\assign{\lb{\resvar}}{\min(\lb{\resvar}, \resvar)}$).
We can then translate lower-bound assertions $\lbassert{\expr}{\resvar}$ in the analogous manner and extend the selectively-amortized resource bounding invariant $\samevarstorerel$ for lower bounds.

\section{Selecting a Decomposition}
\label{sec:selecting-decomposition}

In this section, we describe a way to select amortization groups (\ie a
resource decomposition $\decomp$) and amortization segments (\ie insertions of
amortization $\resetkw$s) to algorithmically realize selectively-amortized
resource bounding. 
As alluded to in \secref{overview}, there is a tension between creating as many
groups and as short segments as possible to focus amortized reasoning only where
it is needed, simplifying the invariant inference needed to do so, versus
not creating too many groups or too short segments that the needed amortization for
precision is lost. 
More specifically, the built-in multiplication $\cnt{\resvar} \cdot
\ub{\resvar}$ we apply for worst-case reasoning around segments simplifies the
necessary invariants needed to prove bounds but only if $\ub{\resvar}$ is
sufficiently precise bound on resource usage per segment.

As hinted at in \secref{decomposing}, the space of possible selective amortizations is huge.
Even with some basic restrictions to make this search more feasible, the remaining space of selective amortizations is still large.
In the remainder of this section, we first characterize when the resource-usage summary $\ub{\resvar}$ is precise based on a notion of \emph{non-interfering amortization segments}.
Then, we describe the basic restrictions and their motivations to use segment non-interference to search within this restricted space.

\paragraph{Non-Interfering Amortization Segments.}

Recall the selective-amortization assertion $\orig{\resvar} \leq \cnt{\resvar} \cdot \ub{\resvar} + \resvar$ and the $\assign{\ub{\resvar}}{\max(\ub\resvar,\resvar)}$ update for a $\reset{\resvar}$ from \secref{resetting}.
We can see that the difference between the sides of the inequality (\ie $(\cnt{\resvar} \cdot \ub{\resvar} + \resvar) - \orig{\resvar}$) comes from a difference between the current upper-bound summary $\ub\resvar$ and the current resource accumulation in $\resvar$ (\ie $\ub\resvar - \resvar$) on a $\reset{\resvar}$.
Thus intuitively, we want to insert amortization resets $\reset{\resvar}$ at locations that would minimize this difference $\ub\resvar - \resvar$ across all such amortization segments.
This observation suggests a definition for segment non-interference:
\begin{definition}[Amortization Segment Non-Interference]\label{def:segment-noninterference}\em
Consider two paths $\ptrace\colon \traceconscmd{\traceconsstate{(\pstoresplit{\lo\pstore}{\hi\pstore})}{\reset{\resvar}} \cdots}{\pstore}{\reset{\resvar}}$ and $\ptrace'\colon \traceconscmd{\traceconsstate{(\pstoresplit{\lo\pstore}{\hi\pstore'})}{\reset{\resvar}} \cdots}{\pstore'}{\reset{\resvar}}$ such that $\Domain(\hi\pstore) = \Domain(\hi\pstore')$.
That is, we consider two amortization segments (\ie paths that start and end in a $\reset{\resvar}$) and partition the input into low variables (\ie $\Domain(\lo\pstore)$) and high variables (\ie $\Domain(\hi\pstore)$).
Then, we say segments $\ptrace$ and $\ptrace'$ are \emph{non-interfering} iff
for any (high) stores $\hi\pstore$ and $\hi\pstore'$, and for any (low) store $\lo\pstore$, we have that $\pstore(\resvar) = \pstore'(\resvar)$.
\end{definition}
We see that if all pairs of amortization segments are non-interfering for a
suitable partition of variables between high and low variables, then the selective amortization is as precise as the fully amortized solution.
Then, we want to balance making amortization segments as small as possible (in order to simplify invariant inference and maximize worst-case reasoning) with the smallest set of low input variables (to maximize non-interference).


\paragraph{Computed Input-Independent Groups and Single Location-Based Segments.}

\Defref{segment-noninterference} suggests an approach to selecting amortization groups and segments if we fix some basic restrictions:
\begin{inparaenum}[(1)]
\item
First, we consider syntactic decomposition transformations $\hypdecomp{\orig\prog}{\prog}$ from the original program $\orig\prog$.
\item
Second, we consider a single insertion of $\reset{\resvar}$ into the transformed program $\prog$ that control-dominates all uses $\auxupdate{\resvar}{\expr}$ for every resource $\resvar$.
Picking a control-dominating location $\loc$ ensures we do not have any $\auxupdate{\resvar}{\expr}$ before a $\reset{\resvar}$, and performing single insertion means we only need to consider segments that start and end at \emph{single location} $\loc$ (where $\mktrans{\loc}{\reset{\resvar}}{\loc'} \in \prog$).
\item\label{restriction-highlow}
Third, we fix the low variables in any segmentation we consider to be the internal variables on which the $\auxupdatekw$s data-depends, leaving any remaining variables at the segment start location $\loc$ to be high, including the inputs to the entry location of the original program $\orig\prog$.
Intuitively, we assume that $\auxupdatekw$s that share computation over internal, low variables are related for amortization.
\end{inparaenum}
However, there is still significant flexibility in choosing the resource decomposition $\decomp$ that defines the amortization groups and the $\auxupdatekw$s-dominating location $\ell$ for each resource $\resvar$ in the transformed program $\prog$---it does not have to be the immediate dominator of the $\auxupdatekw$s.

As we want to create more groups to simplify invariant inference, let us first consider the resource decomposition $\decomp$ such that each syntactic $\auxupdate{\resvar}{\expr}$ in $\orig\prog$ is translated to a unique resource variable and thus placed in a distinct group (\ie such that $|\decomp(\resvar)| = |\SetST{ (\loc,\expr,\loc') }{ \mktrans{\loc}{\auxupdate{\resvar}{\expr}}{\loc'} \in \orig\prog }|$).
However, to find cases where distinct groups are potentially insufficient, we consider possibly merging $\auxupdatekw$ sites pairwise (\ie $\mktrans{\loc_1}{\auxupdate{\resvar_1}{\expr_1}}{\loc_1'}$ and $\mktrans{\loc_2}{\auxupdate{\resvar_2}{\expr_2}}{\loc_2'}$ in the transformed program $\prog$).
Suppose we were to merge groups $\resvar_1$ and $\resvar_2$, then let us consider the immediate dominator $\loc$ of locations $\loc_1$ and $\loc_2$, which defines the possible amortization segments starting from and ending at location $\loc$.
Considering this potential segmentation and the shared low input variables that may affect the value of both $\resvar_1$ and $\resvar_2$ and if the values of these low input variables may change in the segment, then we want to merge these groups based on restriction (\ref{restriction-highlow}) above (otherwise, they are \emph{computed input independent}).
We can then approximate this criteria with standard, backwards data-dependency slices~\cite{DBLP:journals/tse/Weiser84} from the uses $\auxupdate{\resvar_1}{\expr_1}$ and $\auxupdate{\resvar_2}{\expr_2}$.

Once we have fixed a resource decomposition $\decomp$ defining amortization groups, selecting a location $\loc$ to insert each $\reset{\resvar}$ for each $\resvar$ in the transformed program $\prog$ is fairly straightforward.
Following segment non-interference, for any $\auxupdatekw$ sites sharing the same resource $\resvar$ (\ie $L = \SetST{ \loc }{ \mktrans{\loc}{\auxupdate{\resvar}{\expr}}{\loc'} \in \prog }$), find the most immediate dominator of $L$ where we can prove that the low input variables are constant (\ie call this $\auxupdatekw$-dominating location $\loc$, then we have that $\pstore(\lo\var) = \valn$ for some $\valn$, for all low input variables $\lo\var$, in all reachable states $\locstore{\loc}{\pstore}$).
If we can prove that the low input variables are constant in the program up to the amortization segment entry location $\loc$, then we satisfy segment non-interference (up to non-determinism within segments).

Note that because of the tension between precision from simplifying invariant inference versus from amortization, selecting a decomposition is necessarily heuristic. \Secref{decomposing} shows that picking any decomposition is sound, and \secref{experiment} offers evidence that the principled heuristic described here provides a benefit.

\section{Empirical Evaluation}
\label{sec:experiment}

Selective amortization represents a large space of possible approaches between
worst-case and fully amortized reasoning. 
Here we attempt to provide evidence that selective amortization provides a benefit when compared with the two extremes, even with simply the heuristic decomposition strategy described in \secref{selecting-decomposition}.
It is this specific selective amortization strategy that we consider here in our experiments.
We consider the following research question on \emph{Effectiveness}:
Can selective amortization improve the number of
    verified programs when compared with the worst-case and fully-amortized extremes? 

\begin{table}[b]
\footnotesize
\caption{
  Verifying with worst-case (Wor), fully-amortized (Ful), and selectively-amortized (Sel) with two sets of assertions: the most precise bounds and constant-weakened ones. For each configuration, we give the number of assertions proven (n) and the total verification time in seconds (s).
}
\begin{tabular*}{\linewidth}{@{\extracolsep{\fill}}>{\scriptsize}l >{\scriptsize}r >{\scriptsize}r rr rr rr rr rr rr@{}}\toprule
        &       &   & \multicolumn{6}{c}{Most Precise Bounds}      & \multicolumn{6}{c}{Constant-Weakened Bounds} \\ \cmidrule(lr){4-9} \cmidrule(lr){10-15}
        & &       & \multicolumn{2}{c}{Wor} & \multicolumn{2}{c}{Ful} & \multicolumn{2}{c}{Sel} & \multicolumn{2}{c}{Wor} & \multicolumn{2}{c}{Ful} & \multicolumn{2}{c}{Sel}
        \\
        \cmidrule(lr){4-5}
        \cmidrule(lr){6-7}
        \cmidrule(lr){8-9}
        \cmidrule(lr){10-11}
        \cmidrule(lr){12-13}
        \cmidrule(lr){14-15}
  category & num & loc & \multicolumn{1}{l}{(n)} & (s)   & \multicolumn{1}{l}{(n)} & (s)   & \multicolumn{1}{l}{(n)} & (s)   & \multicolumn{1}{l}{(n)} & (s)   & \multicolumn{1}{l}{(n)} & (s)   & \multicolumn{1}{l}{(n)} & (s) \\
  \midrule
  lang3 & 20    & 667   & \textbf{12} & 175.7 & 8     & 44.2  & \textbf{12} & 249.5 & \textbf{14} & 302.7 & \textbf{14} & 89.0    & \textbf{14} & 252.0 \\
  stringutils & 10    & 390   & 2     & 12.9  & \textbf{4} & 196.5 & \textbf{4} & 176.2 & 4     & 101.3 & 5     & 209.0   & \textbf{6} & 264.3 \\
  guava & 3     & 90    & 0     & 0\phantom{.0}     & \textbf{1} & 7.6   & 0     & 0\phantom{.0}     & 2     & 18.3  & \textbf{3} & 30.0    & \textbf{3} & 73.3 \\
  stac  & 3     & 122   & 2     & 118.6 & 2     & 23    & \textbf{3} & 101.1 & 2     & 126.6 & 2     & 22.5  & \textbf{3} & 105.2 \\
  generated & 200   & 3633  & 139   & 1510.2 & 43    & 198.3 & \textbf{175} & 1779.5 & 140   & 1567.8 & 69    & 325.3 & \textbf{180} & 1852.2 \\
  \midrule
  total & 236   & 4902  & 155   & 1817.4 & 58    & 469.6 & \textbf{194} & 2306.3 & 162   & 2116.7 & 93    & 675.8 & \textbf{206} & 2547.0 \\
  \bottomrule
\end{tabular*}%
\label{tbl:experiments}
\end{table}

\paragraph{Effectiveness.}
In \autoref{tbl:experiments}, we summarize the comparison between selective amortization and the two extremes
with the most precise configuration in each category \textbf{bolded}.
For each category, we list the number of programs (num) and the total lines of code (loc).
To test the effect of slightly weaker bound assertions, we consider two sets of assertions: for the most precise bounds and by relaxing the constant coefficients from the most precise bounds.
For each configuration, we applied the same verification tools after transformation with our tool $\toolname$~\cite{zenodo-artifact} implemented in 6,000 lines of Scala,
using Z3~\cite{DBLP:conf/tacas/MouraB08} for SMT solving and ICRA~\cite{DBLP:journals/pacmpl/KincaidBCR19} as an off-the-shelf invariant generator.
%
For the two sets of bound assertions, 194 and 206 programs, respectively, were verified for the selective amortization configuration---more than the number with either extreme.
The improvement over worst-case reasoning comes from amortizing the costs over multiple commands, while
the improvement over fully-amortized reasoning comes from amortizing the costs over subprograms that are smaller than the whole program, so that inferring invariants becomes more manageable for ICRA.


The verification time in \autoref{tbl:experiments} consists of selecting amortizations, realizing amortizations via program transformations, and verifying bound assertions on the transformed programs, which include invariant generation.
We observed that
selecting amortizations and realizing them via program transformations consumed negligible amounts of time;
invariant generation took up more than 95\% of the total time.
Selecting amortizations based on the approach described in \secref{selecting-decomposition} is fast because the selection only requires simple data- and control-dependency analysis.

As noted above, these experiments consider the specific decomposition strategy described in \secref{selecting-decomposition} \emph{on the original benchmarks}, even though we show in \secref{decomposing} that picking any decomposition is sound.
But as alluded to in \secref{decomposing}, programs can be transformed in semantics-preserving ways that then expose different possible decompositions (to either the strategy described in \secref{selecting-decomposition} or even some other one).
Others have made similar observations; for example,
semantic program transformations that split a loop into multiple phases~\cite{DBLP:conf/cav/SharmaDDA11} may simplify the invariant generation by reducing the need for disjunctive invariants.
Indeed, it may strike the best balance between scalability and precision if we can effectively perform different semantic transformations based on the precision we need for proving some desired bounds.


\paragraph{Benchmarks.}
We developed this benchmark suite specifically for the resource bounding problem (as it differs from, for example, the loop bounding problem).
In particular, we collected code from 36 real-world programs (from 4 libraries or suites) that use \Code{StringBuilder}.
Furthermore, we created a suite of 200 synthetic programs generated by randomly nesting and sequencing two common loop idioms that are extracted from actual Java programs.

\section{Related Work}
\label{sec:related-works}

\paragraph{Loop Bound Analysis and Worst-Case Reasoning.}
A large body of work has addressed bounding the number of loop iterations in imperative numeric programs~\cite{DBLP:conf/pldi/GulwaniJK09,DBLP:conf/popl/GulwaniMC09,DBLP:conf/pldi/GulwaniZ10,DBLP:conf/tacas/BrockschmidtEFFG14,DBLP:conf/cav/SinnZV14,DBLP:journals/jar/SinnZV17,DBLP:conf/sas/ZulegerGSV11,DBLP:conf/pldi/Carbonneaux0S15,DBLP:conf/cav/Carbonneaux0RS17}.
These techniques rely on ranking functions to quantitatively track the changes in the rankings of states.
The loop bounding problem can be seen as a special case of the resource bounding problem where the resource of interest is loop iteration, and the cost of each ``use'' (\ie iteration) is a constant 1. Or in other words, the loop bounding problem can be extended to address the resource bounding problem, if one adopts what we call worst-case reasoning to fix a constant upper bound for each resource use. There are works that essentially take this perspective to apply loop bound analysis for invariant inference~\cite{DBLP:journals/jar/SinnZV17,DBLP:conf/tacas/BrockschmidtEFFG14,DBLP:conf/fmcad/CadekDSZ18}.

Worst-case execution time (WCET) analysis~\cite{wilhelm2008worst} is an area of study
that attempts to automatically infer time bounds for machine code, considering
precise models of hardware architectures. It can be seen as another instance of
worst-case reasoning, focusing on defining precise worst-case bound models for 
instructions but generally assuming loop bounds are given or easy to derive.

Our approach is partly inspired by \citet{DBLP:conf/popl/GulwaniMC09} that describes a loop bound analysis because we also rely on a program transformation to simplify the forms of the needed inductive invariants.
At the same time, we improve on this work by
first generalizing the reasoning of loop iterations to general resources, which can change in a non-trivial (\ie non-monotonic and non-constant) way, and then introduce \emph{selective amortization} that mixes in amortized reasoning from the next category of papers to address these challenges.

\paragraph{(Fully-)Amortized Reasoning.}
Several lines of work employ a number of different techniques to precisely reason about resource usage over \emph{full} executions (\ie attempt to perform fully-amortized reasoning).
The COSTA project~\cite{DBLP:conf/esop/AlbertAGPZ07,DBLP:conf/iwmm/AlbertGG07,DBLP:conf/iwmm/AlbertGG09,DBLP:conf/vmcai/AlbertGM11,DBLP:conf/sas/Alonso-BlasG12}, which adopts the recurrence relation approach, reasons about resource usage by first abstracting program semantics into a set of recurrence relations and then finding closed-form solutions to these recurrence relations.
The RAML project~\cite{DBLP:conf/popl/HoffmannAH11,DBLP:conf/popl/HoffmannDW17} analyzes the resource usage of functional programs with the potential method.
This approach encodes the changes of a potential
with linear programming constraints over the unknown coefficients of pre-determined bound templates.
\citet{DBLP:conf/pldi/Carbonneaux0S15,DBLP:conf/cav/Carbonneaux0RS17} adapts this approach to numerical imperative programs.
\citet{DBLP:conf/esop/Atkey10} (and improvements~\cite{DBLP:conf/esop/GueneauCP18,DBLP:conf/itp/GueneauJCP19}) develop expressive program logics that extend type-based amortized resource analysis with resource reasoning over heap data structures.
The above approaches can be viewed as instances of fully-amortized reasoning, because it is in an amortized manner that they encode the sum of the resource usage into systems of constraints~\cite{DBLP:conf/esop/AlbertAGPZ07,DBLP:conf/iwmm/AlbertGG07,DBLP:conf/iwmm/AlbertGG09,DBLP:conf/vmcai/AlbertGM11,DBLP:conf/sas/Alonso-BlasG12,DBLP:conf/pldi/Carbonneaux0S15,DBLP:conf/cav/Carbonneaux0RS17} or perform deductive proofs that amortize costs~\cite{DBLP:conf/esop/Atkey10,DBLP:conf/esop/GueneauCP18,DBLP:conf/itp/GueneauJCP19}.
The key challenge in fully-amortized reasoning is to infer complex inductive invariants, which are the solutions of the constraint systems in, for example, COSTA and RAML.
Instead, our approach may simplify the forms of the required invariants by decomposing the resource usage into groups and segments of amortized costs.
Since our approach is agnostic to the underlying amortized reasoning engine, any fully-amortized reasoning approach, such as the above ones, can potentially be used in place of the relational inductive invariant generator applied in this paper.
\section{Conclusion}
\label{sec:conclusion}

In this paper we address the problem of automatically proving resource bounds, where resource usage is expressed via an integer-typed variable.
We present a framework for selectively-amortized reasoning that mixes worst-case and fully amortized reasoning via a property decomposition and a program transformation.
We show that proving bounds in any such decomposition yields a sound resource bound in the original program, and we give an algorithm for selecting an effective decomposition.
Our empirical evaluation provides evidence that selective amortization effectively leverages both worst-case and amortized reasoning.


\subsubsection*{Acknowledgements}

We thank Pavol \v{C}ern\'{y} for his valuable contributions in the early stages of this research.
We also thank the anonymous reviewers and members of the CUPLV lab for their helpful reviews and suggestions.
This research was supported in part by the Defense Advanced Research Projects Agency under grant FA8750-15-2-0096, and also by the National Science Foundation under grant CCF-2008369.

%
%
%
\bibliographystyle{plainnat}
\bibliography{conference.short,00-references.short}


\ifTR
\appendix





\section{Soundness of Selectively-Amortized Resource Bounding}
\label{sec:proofs}

In this section, we prove the soundness of using any selective amortization decomposition to approximate resource usage, which is stated as \autoref{thm:soundness} in \autoref{sec:resetting-soundness}.

\subsection{Soundness of Selectively-Amortized Resource Bounding for Commands}

\begin{figure}[tb]
\begin{mathpar}
\fbox{\evalcmd{\pstore}{\cmd}{\pstore'}} \\

\infer[E-Skip]{
}{
  \evalcmd{\pstore}{ \cmdskip }{ \pstore }
}

\infer[E-Assign]{
  \evalexpr{\pstore}{\expr}{\val}
}{
  \evalcmd{\pstore}{ \assign{\var}{\expr} }{ \mapext{\pstore}{\var}{\val} }
}

\infer[E-Assume]{
  \evalexpr{\pstore}{\expr}{\kwtrue}
}{
  \evalcmd{\pstore}{ \assume{\expr} }{ \pstore }
}
\end{mathpar}
\caption{Standard evaluation semantics for the core imperative commands $\cmdskip$, $\assign{\var}{\expr}$, and $\assume{\expr}$.}
\label{fig:corelang-standardcmds}
\end{figure}
In \figref{corelang-standardcmds}, we give the inference rules defining the standard evaluation semantics for the core imperative commands elided in \secref{decomposing}. Also recall from \secref{decomposing} that we consider a standard expression language $\expr$ that we leave unspecified, aside from including program variables $\var$ and its value forms $\val$ having integers $\valn$ and booleans $\valb$. But note that expressions do not include resources $\resvar$.

Recall that selectively-amortized conformance relation $\samevarstore{\pstore}{\pstore'}$ states that stores $\pstore$ and $\pstore'$ are equal on program variables $\Vars(\pstore)$ and may differ only in resources.
Thus, we assume that expressions in conforming stores evaluate to the same values.
\begin{property}[Expression Evaluation under Conforming Stores]\label{prp:expevalunderconforming}
If $\samevarstore{\orig\pstore}{\pstore}$, then for any expression $\expr$ and value $\val$,
$\evalexpr{\orig\pstore}{\expr}{\val}$ iff $\evalexpr{\pstore}{\expr}{\val}$.
\end{property}

We now prove the soundness of selectively-amortized resource bounding for commands (i.e., part \ref{soundness-cmds} of \autoref{thm:soundness}), which we restate as \autoref{thm:sound-cmd}. The essence of this statement is that for any command decomposition $\hypdecomp{\orig\cmd}{\cmd}$, if the transformed command $\cmd$ makes progress, then the original command $\orig\cmd$ can also make progress preserving the selectively-amortized resource bounding invariant $\samevarstorerel$. Note that the transformed command is allowed to get stuck when the original command would have made progress, corresponding to a sound but imprecise selective amortization.

\begin{lemma}[Soundness of Selectively-Amortized Resource Bounding for Commands]\label{thm:sound-cmd}
  If $\hypdecomp{\orig\cmd}{\cmd}$ and
  $\evalcmd{\pstore}{\cmd}{\pstore'}$, and
  $\samevarstore{\orig\pstore}{\pstore}$, then
  $\evalcmd{\orig\pstore}{\orig\cmd}{\orig{\pstore'}}$ with $\samevarstore{\orig\pstore'}{\pstore'}$.
\end{lemma}
\begin{proof}
By cases on the structure of the derivation $\fsD$ of $\hypdecomp{\orig\cmd}{\cmd}$. Assume we have a derivation $\fsE$ of $\evalcmd{\pstore}{ \cmd}{\pstore'}$
such that $\samevarstore{\orig\pstore}{\pstore}$, we want to construct a derivation $\orig\fsE$ of
$\evalcmd{\orig\pstore}{\orig\cmd}{\orig{\pstore'}}$ with $\samevarstore{\orig\pstore'}{\pstore'}$.
\begin{case}
\(
\fsD =
\infer*[right=D-Use]{
  \resvar' \in \resvarseq
}{
  \hypdecomp[\decompcons{\decomp}{\decompto{\resvar}{\resvarseq}}]{ \auxupdate{\resvar}{\expr} }{ \auxupdate{\resvar'}{\expr} }
}
\) \\\\
By cases on command evaluation, we have that
\[
  \fsE = 
  \infer*[right=E-Use]{
    \fsE' ::
    \evalexpr{\pstore}{\expr}{\valn}
  }{
    \evalcmd{\pstore}{ \auxupdate{\resvar'}{\expr} }{ \pstore' }
  }
  \quad\text{where}\quad \pstore' = \mapext{\pstore}{\resvar'}{\pstore(\resvar') + \valn} \;.
\]
By expression evaluation under conforming stores (\prpref{expevalunderconforming}) on $\fsE'$ using
$\samevarstore{\orig\pstore}{\pstore}$, we have a $\orig\fsE' :: \evalexpr{\orig\pstore}{\expr}{\valn}$,
so let
\[
  \orig\fsE \defeq
  \infer*[right=E-Use]{
    \orig\fsE' ::
    \evalexpr{\orig\pstore}{\expr}{\valn}
  }{
    \evalcmd{\orig\pstore}{ \auxupdate{\resvar}{\expr} }{ \orig\pstore' }
  }
  \quad\text{where}\quad \orig\pstore' = \mapext{\orig\pstore}{\resvar}{\orig\pstore(\resvar) + \valn}
\]
Now, we have that
\[ \begin{array}{@{\extracolsep{0.5em}}rcl}
  \orig\pstore'(\resvar) & = & \orig\pstore(\resvar) + \valn
  \\[3ex]
  & \leq &
  \pstore(\cnt{\resvar'}) \cdot \pstore(\ub{\resvar'}) + \pstore(\resvar') + \valn +
  \Sum_{\resvar'' \in \decomp(\resvar) \backslash \Set{\resvar'}}
    \pstore(\cnt{\resvar''}) \cdot \pstore(\ub{\resvar''}) + \pstore(\resvar'')
  \\[3ex]
  & = &
  \Sum_{\resvar'' \in \decomp(\resvar)}
    \pstore'(\cnt{\resvar''}) \cdot \pstore'(\ub{\resvar''}) + \pstore'(\resvar'') \;.
\end{array} \]
So it is the case that $\samevarstore{\orig\pstore'}{\pstore'}$ (as all other resources
$\Domain(\orig\pstore) \backslash \Set{\resvar}$ remain unchanged).
\end{case}

\begin{case}
\(
\fsD =
\infer*[right=D-UBCheck]{
}{
  \hypdecomp[\decompcons{\decomp}{\decompto{\resvar}{\resvarseq}}]{ \ubassert{\resvar}{\expr} }{ \ubassert{\resvarseq}{\expr} }
}
\) \\\\
By cases on command evaluation, we have that
\[
  \fsE = 
\infer*[right=E-UBCheck]{
  \fsE' :: \evalexpr{\pstore}{\expr}{\valn}
  \\
  \left(\Sum_{\resvar \in \resvarseq}
  \pstore(\cnt{\resvar}) \cdot \pstore(\ub{\resvar}) + \pstore(\resvar)\right)
  \leq \valn
}{
  \evalcmd{\pstore}{ \ubassert{\resvarseq}{\expr} }{ \pstore }
}  
\]
By $\samevarstore{\orig\pstore}{\pstore}$, we have that
\[
  \orig\pstore(\resvar) \leq
  \left(\Sum_{\resvar \in \resvarseq}
  \pstore(\cnt{\resvar}) \cdot \pstore(\ub{\resvar}) + \pstore(\resvar)\right)
  \leq \valn
\]
and by expression evaluation under conforming stores (\prpref{expevalunderconforming}) on $\fsE'$ using
$\samevarstore{\orig\pstore}{\pstore}$, we have a $\orig\fsE' :: \evalexpr{\orig\pstore}{\expr}{\valn}$.
So we have $\orig\fsE$ as follows:
\[
  \orig\fsE  \defeq
\infer*[right=E-UBCheck]{
  \orig\fsE' :: \evalexpr{\orig\pstore}{\expr}{\valn}
  \\
  \orig\pstore(\resvar)
  \leq \valn
}{
  \evalcmd{\orig\pstore}{ \ubassert{\resvarseq}{\expr} }{ \orig\pstore }
}  
\]
and $\samevarstore{\orig\pstore}{\pstore}$ by assumption.
\end{case}

\begin{case}
\(
\fsD =
\infer*[right=D-Reset]{
}{
   \hypdecomp{ \cmdskip }{ \reset{\resvar} }
}
\) \\\\
By cases on command evaluation, we have that
\[
  \fsE = 
\infer*[right=E-Reset]{
  \pstore' =
  \mapext{
    \mapext{
      \mapext{
        \pstore}
             {\cnt{\resvar}}{ \pstore(\cnt{\resvar}) + 1}
    }
           {
             \ub{\resvar}}{ \max(\pstore(\ub{\resvar}), \pstore(\resvar))
           }
  }{\resvar}{0}
}{
  \evalcmd{\pstore}{ \reset{\resvar} }{ \pstore' }
}
\]
We construct $\orig\fsE$ as follows:
\[
  \orig\fsE \defeq
\infer*[right=E-Skip]{
}{
  \evalcmd{\orig\pstore}{ \cmdskip }{\orig\pstore}
}
\]
So we wish to show that $\samevarstore{\orig\pstore}{\pstore'}$. To do so, assume
$\orig\resvar$ is a resource such that $\resvar \in \decomp(\orig\resvar)$.
Then, we have that
\[ \begin{array}{rcl}
  \orig\pstore(\orig\resvar)
  & \leq &
  \left(\Sum_{\resvar' \in \decomp(\orig\resvar)}
  \pstore(\cnt{\resvar'}) \cdot \pstore(\ub{\resvar'}) + \pstore(\resvar')\right)
  \\
  & = &
  \pstore(\cnt{\resvar}) \cdot \pstore(\ub{\resvar}) + \pstore(\resvar)
  + \left(\Sum_{\resvar' \in \decomp(\orig\resvar) \backslash \Set{\resvar}}
  \pstore(\cnt{\resvar'}) \cdot \pstore(\ub{\resvar'}) + \pstore(\resvar')\right)
  \\
  & \leq &
  (\pstore(\cnt{\resvar}) + 1) \cdot \max(\pstore(\ub{\resvar}),\pstore(\resvar))
  + \left(\Sum_{\resvar' \in \decomp(\orig\resvar) \backslash \Set{\resvar}}
  \pstore(\cnt{\resvar'}) \cdot \pstore(\ub{\resvar'}) + \pstore(\resvar')\right)
  \\
  & = &
  \pstore'(\cnt{\resvar}) \cdot \pstore'(\ub{\resvar}) + \pstore'(\resvar)
  + \left(\Sum_{\resvar' \in \decomp(\orig\resvar) \backslash \Set{\resvar}}
  \pstore'(\cnt{\resvar'}) \cdot \pstore'(\ub{\resvar'}) + \pstore'(\resvar')\right)
  \\
  & = &
  \left(\Sum_{\resvar' \in \decomp(\orig\resvar)}
  \pstore'(\cnt{\resvar'}) \cdot \pstore'(\ub{\resvar'}) + \pstore'(\resvar')\right)
\end{array}\]
Thus, we have $\samevarstore{\orig\pstore}{\pstore'}$.
\end{case}

\begin{case}
\(
\fsD =
\infer*[right=D-Command]{
  \cmd \in \Set{ \cmdskip, \assign{\var}{\expr}, \assume{\expr}}
}{
  \hypdecomp{ \cmd }{ \cmd }
}
\) \\\\
Consider the cases on the command $\cmd$.
\begin{subcase}
\( \cmd = \cmdskip \). By cases on command evaluation, we have that $\evalcmd{\pstore}{ \cmdskip }{ \pstore }$,
so we simply construct
\[
 \orig\fsE \defeq
 \infer*[right=E-Skip]{
 }{
   \evalcmd{\orig\pstore}{ \cmdskip }{ \orig\pstore }
 }
\]
and $\samevarstore{\orig\pstore}{\pstore}$ by assumption.
\end{subcase}
\begin{subcase}
\( \cmd = \assign{\var}{\expr} \).
By cases on command evaluation, we have that
\[
  \fsE = 
  \infer*[right=E-Assign]{
    \fsE' ::
    \evalexpr{\pstore}{\expr}{\val}
  }{
    \evalcmd{\pstore}{ \assign{\var}{\expr} }{ \pstore' }
  }
  \quad\text{where}\quad \pstore' = \mapext{\pstore}{\var}{\val} \;.
\]
By expression evaluation under conforming stores (\prpref{expevalunderconforming}) on $\fsE'$ using $\samevarstore{\orig\pstore}{\pstore}$,
we have that $\orig\fsE' :: \evalexpr{\orig\pstore}{\expr}{\val}$, so let
\[
  \orig\fsE \defeq
  \infer*[right=E-Assign]{
    \orig\fsE' ::
    \evalexpr{\orig\pstore}{\expr}{\val}
  }{
    \evalcmd{\orig\pstore}{ \assign{\var}{\expr} }{ \orig\pstore' }
  }
  \quad\text{where}\quad \orig\pstore' = \mapext{\orig\pstore}{\var}{\val} \;.
\]
Now, we have that $\orig\pstore(\var) = \pstore(\var)$, so it is the case that
$\samevarstore{\orig\pstore'}{\pstore'}$ (as all other program variables
$\Vars(\orig\pstore) \backslash \Set{\var}$ remain unchanged).
\end{subcase}
\begin{subcase}
\( \cmd = \assume{\expr} \).
By cases on command evaluation, we have that
\[
  \fsE = 
  \infer*[right=E-Assume]{
    \fsE' ::
    \evalexpr{\pstore}{\expr}{\kwtrue}
  }{
    \evalcmd{\pstore}{ \assume{\expr} }{ \pstore }
  }
\]
By expression evaluation under conforming stores (\prpref{expevalunderconforming}) on $\fsE'$ using $\samevarstore{\orig\pstore}{\pstore}$,
we have that $\orig\fsE' :: \evalexpr{\orig\pstore}{\expr}{\kwtrue}$, so let
\[
  \orig\fsE \defeq
  \infer*[right=E-Assume]{
    \orig\fsE' ::
    \evalexpr{\orig\pstore}{\expr}{\kwtrue}
  }{
    \evalcmd{\orig\pstore}{ \assume{\expr} }{ \orig\pstore }
  }
\]
and $\samevarstore{\orig\pstore}{\pstore}$ by assumption.
\end{subcase}
\end{case}
\qed
\end{proof}

\subsection{Soundness of Selectively-Amortized Resource Bounding for Paths}



We now prove soundness selectively-amortized resource bounding for paths (i.e., part \ref{soundness-paths} of \autoref{thm:soundness}), which we restate as \autoref{thm:sound-path}. The statement says that for any path decomposition
$\hypdecomp{\orig\ptrace}{\ptrace}$, if the transformed path $\ptrace$ does not get stuck, then the original path $\orig\ptrace$ also does not.

Intuitively, this holds because both the transformed path $\ptrace$ and the original path $\orig\ptrace$ start in the same initial state. Note that we assume the initial state has no resource use; that is, resource variables $\resvar$ are initialized as described in \secref{resetting} (i.e., $\resvar = 0$, $\ub{\resvar} = 0$, $\cnt{\resvar} = -1$ for any resource $\resvar$). Then at each step, the command decomposition ensures that the selectively-amortized resource-bounding invariant $\samevarstorerel$ is preserved.

\begin{lemma}[Soundness of Selectively-Amortized Resource Bounding for Paths]\label{thm:sound-path}
  If $\hypdecomp{\orig\ptrace}{\ptrace}$ and $\ptraceok{\ptrace}$,
  then $\ptraceok{\orig\ptrace}$.
\end{lemma}
\begin{proof}
By induction on the structure of the derivation $\fsO$ of $\ptraceok{\ptrace}$.
Assume we have a derivation $\fsD$ of $\hypdecomp{\orig\ptrace}{\ptrace}$ to construct a derivation $\orig\fsO$ of $\ptraceok{\orig\ptrace}$.
\begin{case}
\(
\fsO =
\infer*[right=Ok-Init]{
}{
  \ptraceok{\pstate}
}
\) \\\\
By cases on the selectively-amortized resource bounding on paths, we have that derivation 
\[ \fsD =
\infer*[right=D-Init]{
}{
  \hypdecomp{ \pstate }{ \pstate }
} \]
So by assumption, $\ptraceok{\pstate}$ (i.e., $\orig\ptrace = \ptrace = \pstate$).
\end{case}
\begin{case}
\(
\fsO =
\infer*[right=Ok-Step]{
  \fsO' :: \ptraceok{ \traceconsstate{\ptrace'}{\locstore{\loc}{\pstore}} } \\
  \fsE :: \evalcmd{\pstore}{\cmd}{\pstore'}
}{
  \ptraceok{
    \traceconsstate{\traceconscmd{\ptrace'}{\pstore}{\cmd}}{\locstore{\loc'}{\pstore'}} 
  }
}
\) \\\\
By cases on the selectively-amortized resource bounding on paths, we have that derivation $\fsD ::
\hypdecomp{
  \traceconsstate{\traceconscmd{\orig\ptrace'}{\orig\pstore}{\orig\cmd}}{\locstore{\orig\loc'}{\orig\pstore'}} 
}{
  \traceconsstate{\traceconscmd{\ptrace'}{\pstore}{\cmd}}{\locstore{\loc'}{\pstore'}} 
}$
for some $\orig\ptrace'$, $\orig\pstore$, $\orig\cmd$, $\orig\loc'$, and $\orig\pstore'$.
And by inversion on $\fsD$, we have $\fsD' ::
\hypdecomp{
  \traceconscmd{\orig\ptrace'}{\orig\pstore}{\orig\cmd}
}{
  \traceconscmd{\ptrace'}{\pstore}{\cmd}
}$.
By cases on the selectively-amortized resource bounding on paths, we have that derivation $\fsD'$ is as follows:
\[
\fsD' =
\infer*[right=D-AppendCommand]{
  \fsD'' :: \hypdecomp{
    \traceconsstate{\orig\ptrace'}{\locstore{\orig\ell}{\orig\pstore}}
  }{
    \traceconsstate{\ptrace'}{\locstore{\ell}{\pstore}}
  }
  \\
  \fsC :: \hypdecomp{\orig\cmd}{\cmd}
}{
  \hypdecomp{
    \traceconscmd{\orig\ptrace'}{\orig\pstore}{\orig\cmd}
   }{
    \traceconscmd{\ptrace'}{\pstore}{\cmd}
   }
}
\]
By the i.h. on $\fsO'$ with $\fsD''$, we have a derivation $\orig\fsO'$ of $\ptraceok{ \traceconsstate{\orig\ptrace'}{\locstore{\orig\loc}{\orig\pstore}} }$.

We now show that there is a derivation $\orig\fsE$ of $\evalcmd{\orig\pstore}{\orig\cmd}{\orig\pstore'}$. By cases on the selectively-amortized resource bounding on paths, we have two cases for $\fsD''$.
\begin{subcase}
\(
\fsD'' =
\infer*[right=D-Step]{
  \hypdecomp{\orig\ptrace'}{\ptrace'}
  \\
  \ptraceok{
    \traceconsstate{\ptrace'}{\locstore{\ell}{\pstore}}
  }
  \\
  \samevarstore{
    \locstore{\orig\ell}{\orig\pstore}
  }{
    \locstore{\ell}{\pstore}
  }
}{
  \hypdecomp{
    \traceconsstate{\ptrace}{\locstore{\orig\ell}{\orig\pstore}}
   }{
    \traceconsstate{\ptrace'}{\locstore{\ell}{\pstore}}
   }
}
\
\) \\\\
So we have that $\samevarstore{\orig\pstore}{\pstore}$.
\end{subcase}
\begin{subcase}
\(
\fsD'' =
\infer*[right=D-Init]{
}{
  \hypdecomp{
    \locstore{\ell}{\pstore}
  }{
    \locstore{\ell}{\pstore}
  }
}
\quad\text{(where there is no $\resvar$ in $\Domain(\pstore)$)}
\) \\\\
Note that with no resources $\resvar$ in $\pstore$, we have that $\samevarstore{\pstore}{\pstore}$.
\end{subcase}
So in both subcases, we have that $\samevarstore{\orig\pstore}{\pstore}$. By the soundness of selectively-amortized resource bounding for commands (\autoref{thm:sound-cmd}) with $\fsC$, $\fsE$, and $\samevarstore{\orig\pstore}{\pstore}$, we have a derivation $\orig\fsE$ of $\evalcmd{\orig\pstore}{\orig\cmd}{\orig\pstore'}$.

Finally, we construct $\orig\fsO$ as follows:
\[
\orig\fsO \defeq
\infer*[right=Ok-Step]{
  \orig\fsO' :: \ptraceok{ \traceconsstate{\orig\ptrace'}{\locstore{\orig\loc}{\orig\pstore}} } \\
  \orig\fsE :: \evalcmd{\orig\pstore}{\orig\cmd}{\orig\pstore'}
}{
  \ptraceok{
    \traceconsstate{\traceconscmd{\orig\ptrace'}{\orig\pstore}{\orig\cmd}}{\locstore{\orig\loc'}{\orig\pstore'}} 
  }
}
\]
\end{case}
\qed
\end{proof}

\subsection{Soundness of Selectively-Amortized Resource Bounding for Programs}

Finally, we show the soundness of selectively-amortized resource bounding for programs (i.e., part \ref{soundness-programs} of \autoref{thm:soundness}), which we restate as \autoref{thm:sound-prog}. Note that any path through a program is well-formed (i.e., it is straightforward to show that if $\ptrace \in \denote{\prog} \pstate$, then $\ptraceok{\ptrace}$).
\begin{lemma}[Soundness of Selectively-Amortized Resource Bounding for Programs]\label{thm:sound-prog}
  If $\hypdecomp{\orig\prog}{\prog}$
      and $\ptrace \in \denote{\prog} \pstate$,
      then there is a $\orig\ptrace \in \denote{\orig\prog} \pstate$
      s.t. $\hypdecomp{\orig\ptrace}{\ptrace}$.
\end{lemma}
\begin{proof}
By induction on the structure of $\denote{\prog} \pstate$. Assume $\hypdecomp{\orig\prog}{\prog}$ to construct a derivation
$\fsD$ of $\hypdecomp{\orig\ptrace}{\ptrace}$ for some $\orig\ptrace \in \denote{\orig\prog} \pstate$.
\begin{case}$\ptrace = \pstate$. Trivially, $\pstate \in \denote{\orig\prog}\pstate$ and let
\[
\fsD \defeq
\infer*[right=D-Init]{
}{
  \hypdecomp{\pstate}{\pstate}
}
\]
\end{case}
\begin{case}$\ptrace =
\traceconsstate{\traceconscmd{\ptrace'}{\pstore}{\cmd}}{\locstore{\loc'}{\pstore'}}$ where
$\traceconsstate{\ptrace'}{\locstore{\ell}{\pstore}} \in \denote{\prog} \pstate$ such that
$\stepprog{ \locstore{\loc}{\pstore} }{ \locstore{\loc'}{\pstore'} }$ (for some $\ell$).
By the i.h. on
$\traceconsstate{\ptrace'}{\locstore{\ell}{\pstore}}$,
there is a 
$\traceconsstate{\orig\ptrace'}{\locstore{\ell}{\orig\pstore}} \in \denote{\orig\prog} \pstate$ such that
$\fsD' :: \hypdecomp{\traceconsstate{\orig\ptrace'}{\locstore{\ell}{\orig\pstore}}}{\traceconsstate{\ptrace'}{\locstore{\ell}{\pstore}}}$.
By inversion on $\fsD'$ (and cases on selectively-amortized resource bounding on paths), we have $\samevarstore{\orig\pstore}{\pstore}$.

By inversion on $\stepprog{ \locstore{\loc}{\pstore} }{ \locstore{\loc'}{\pstore'}  }$, we have that $\mktrans{\loc}{\cmd}{\loc'} \in \prog$ and $\fsE :: \evalcmd{\pstore}{ \cmd }{ \pstore' }$.
Also, because $\hypdecomp{\orig\prog}{\prog}$, we have that $\fsC :: \hypdecomp{\orig\cmd}{\cmd}$ for some $\mktrans{\loc}{\orig\cmd}{\loc'} \in \orig\prog$. Then, by the soundness of selective-amortized resource bounding for commands (\autoref{thm:sound-cmd}) on $\fsC$, $\fsE$, and $\samevarstore{\orig\pstore}{\pstore}$, we have $\orig\fsE :: \evalcmd{\orig\pstore}{ \orig\cmd }{ \orig\pstore' }$.  Thus, we have $\stepprog{ \locstore{\loc}{\orig\pstore} }{ \locstore{\loc'}{\orig\pstore'} }$ and
$\traceconsstate{\traceconscmd{\orig\ptrace'}{\orig\pstore}{\orig\cmd}}{\locstore{\loc'}{\orig\pstore'}} \in \denote{\orig\prog} \pstate$.

We then construct $\fsD$ as follows:
\[
\infer*[right=D-Step]{
  \infer*[Right=D-AppendCommand,vdots=3ex]{
    \fsD' :: \hypdecomp{
      \traceconsstate{\orig\ptrace'}{\locstore{\loc}{\orig\pstore}}
    }{
      \traceconsstate{\ptrace'}{\locstore{\loc}{\pstore}}
    }
    \quad
    \fsC :: \hypdecomp{\orig\cmd}{\cmd}
  }{
    \hypdecomp{
      \traceconscmd{\orig\ptrace'}{\orig\pstore}{\orig\cmd}
     }{
      \traceconscmd{\ptrace'}{\pstore}{\cmd}
     }
  }  
  \quad
  \quad
  \samevarstore{\locstore{\loc'}{\orig\pstore'}}{\locstore{\loc'}{\pstore'}}
}{
  \hypdecomp{
    \traceconsstate{\traceconscmd{\orig\ptrace'}{\orig\pstore}{\orig\cmd}}{\locstore{\loc'}{\orig\pstore'}} 
   }{
    \traceconsstate{\traceconscmd{\ptrace'}{\pstore}{\cmd}}{\locstore{\loc'}{\pstore'}}
   }
}
\]
Note that for presentation, we elide the well-formedness premise of \TirName{D-Step} in the above (i.e., $\ptraceok{
  \traceconsstate{\traceconscmd{\ptrace'}{\pstore}{\cmd}}{\locstore{\loc'}{\pstore'}}
}$), which we have from $\traceconsstate{\traceconscmd{\ptrace'}{\pstore}{\cmd}}{\locstore{\loc'}{\pstore'}} \in \denote{\prog} \pstate$.
\end{case}
\qed
\end{proof}

\fi

\end{document}